\documentclass[aps,pra,twocolumn]{revtex4}
\usepackage[dvips]{epsfig}
\usepackage[usenames]{color}
\usepackage{pstricks}
\usepackage{amsmath}
\usepackage{amssymb}
\usepackage{subfigure}
\usepackage{afterpage}

\begin{document}

\title{Dynamics of ultracold molecules in confined geometry and electric field}

\author{Goulven Qu{\'e}m{\'e}ner and John L. Bohn}
\affiliation{JILA, University of Colorado,
Boulder, C0 80309-0440, USA}

\date{\today}

\begin{abstract}
We present a time-independent quantum formalism
to describe the dynamics of molecules with permanent electric dipole moments
in a two-dimensional confined geometry such as a one-dimensional optical lattice,
in the presence of an electric field.
Bose/Fermi statistics and selection rules play a crucial role in the dynamics.
As examples, we compare the dynamics of confined fermionic
and bosonic polar KRb molecules
under different confinements and electric fields.
We show how chemical reactions can be suppressed,
either by a ``statistical suppression'' which applies for fermions at small electric fields and confinements,
or by a ``potential energy suppression'', which applies for both fermions and bosons at high electric fields and confinements.
We also explore collisions that transfer molecules from one state of the confining potential to another.
Although these collisions can be significant, 
we show that they do not play a role in 
the loss of the total number of molecules in the gas.
\end{abstract}


\maketitle

\font\smallfont=cmr7

\section{Introduction}

Experimental evidence for 
ultracold chemistry
of quantum-state controlled molecules~\cite{Ospelkaus10-SCIENCE}
and dipolar collisions in the quantum regime~\cite{Ni10-NATURE}
has been obtained recently for
fermionic KRb molecules
in the lowest electronic, vibrational, rotational quantum state~\cite{Ni08-SCIENCE}
and well-defined hyperfine states~\cite{Ospelkaus10-PRL}.
Bosonic species of KRb have also been formed recently~\cite{Aikawa10-ARXIV}
as well as other alkali polar molecules such as RbCs~\cite{Sage05} and LiCs~\cite{Deiglmayr08}.
The exoergic reaction KRb + KRb $\to$ K$_2$+ Rb$_2$~\cite{Zuchowski10,Byrd10,Meyer10}
prevents long trap lifetimes of these molecules, especially
in electric fields, where the chemical reactivity
increases as the sixth power of the dipole moment 
induced by the electric field~\cite{Ni10-NATURE,Quemener10-QT}.
Lifetimes are then typically of the order of 10~ms
for experimental electric fields.
However, polar molecules offer 
long range and anisotropic dipolar interactions
in electric fields.
If the molecules are confined in  optical lattices,
they can be stabilize against collisions and chemical reactions~\cite{Buchler07,Micheli07,Ticknor10,Quemener10-2D,Micheli10-PRL,Dincao10},
if the dipoles are polarized in the direction of a tight confinement.
If these molecules are confined into the ground state of 
a realistic one dimensional optical lattice,
electric field suppression of chemical reactions
is expected to occur, yielding lifetimes of KRb molecules of $\simeq$~1~s
and elastic scattering rates 100 times more efficient
than chemical reaction rates~\cite{Quemener10-2D,Micheli10-PRL}.
Both of these are needed
to achieve molecular evaporative cooling
and to reach the quantum regime where the phase-space density is high. 
For fermionic molecules, creation of degenerate Fermi gases of dipoles
will likely be possible. In case of bosonic molecules, Bose--Einstein
condensates can instead be formed.
This will reveal exciting physics with ultracold 
controlled molecules in the quantum regime~\cite{Carr09,Micheli06-NATURE,Demille02,Yelin06}.

We address in this paper two important points regarding collisions in a lattice.
First, suppression of confined chemical reactions in electric fields
can be obtained by using the centrifugal repulsion 
of fermionic molecules in the same internal 
state (electronic, vibrational, rotational and spin) 
and in the same confining state of the one dimensional optical lattice.
The centrifugal repulsion comes from the statistics 
of identical fermions in indistinguishable
states.
This requires only comparatively small dipoles and weak confinements.
Suppression 
that relies directly on the confining potential and the repulsion due to electric dipoles
can also be obtained,
but requires larger dipoles
and stronger confinements.
It does, however, suppress both bosons and fermions,
in indistinguishable states or not, or even for different polar molecules.

Secondly, realistic experimental dynamics of polar molecules
in confined geometry is more complicated
than the ideal case used in the recent
theoretical works~\cite{Quemener10-2D,Micheli10-PRL},
where only molecules in the ground state of the lattice
were considered.
Realistically, molecules can also be formed
in excited states of the optical lattice,
depending for example on the temperature, the strength of the confinement, and the way
the optical lattice is turned on~\cite{Miranda10}.
It is therefore important to know how rapidly 
collisions can populate higher confining states, which could after all,
contribute to re-thermalization; and how do the molecules in these
excited states affect the loss rate of the total molecules.
These questions are important for ongoing experiments of KRb molecules
in an optial lattice~\cite{Miranda10}.

In this article, we extend the formalism developped in our former work~\cite{Quemener10-2D}.
We describe in section II the dynamics of molecules in an arbitrary initial confining state of the lattice,
and consider the possibility for the molecules to leave such a state for another after a collision.
In section III, we show how chemical reaction can be suppressed for fermionic and bosonic KRb molecules
under different confinements and electric fields.
In section IV, we discuss the importance of inelastic collisions
of molecules in different confining states.
Finally, we conclude in section V.

In the following, quantities are expressed in S.I. units, unless explicitly stated otherwise.
Atomic units (a.u.) are obtained by setting $\hbar = 4 \pi \varepsilon_0 = 1$.

\section{Theoretical formalism}

\begin{figure} [t]
\begin{center}
\includegraphics*[bb=80 225 550 900,width=12cm,keepaspectratio=true,angle=-90]{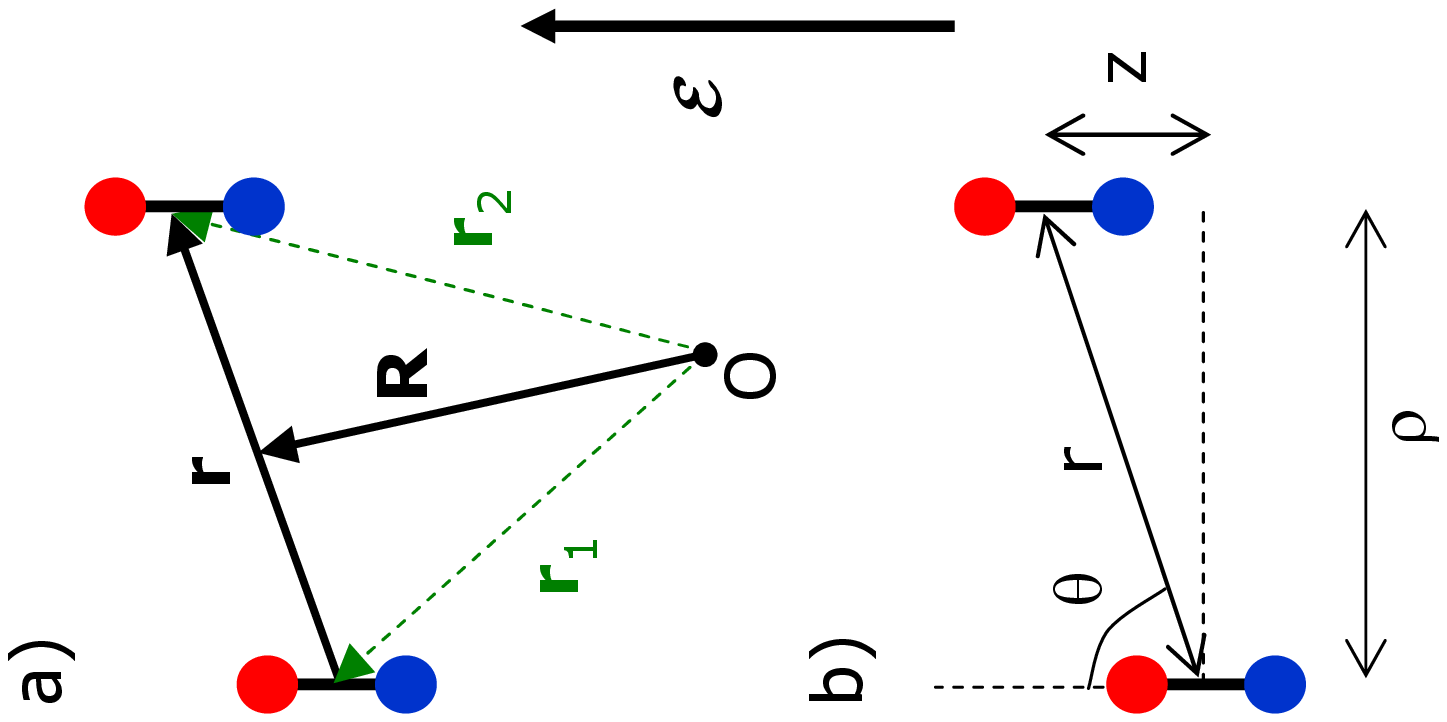}
\caption{(Color online) a) Position vectors of the molecules.
The electric field is along the $z$ direction.
b) Spherical coordinates $(r, \theta)$ and cylindrical coordinates $(\rho, z)$ of the relative coordinate. 
We suppose $\varphi = 0$ in the picture.
\label{COORD-FIG}
}
\end{center}
\end{figure}

In this section, we explain the theoretical formalism we use.
Former studies have dealt with collisions in two dimensions~\cite{Petrov01,Li09,Ticknor10,Quemener10-2D,Micheli10-PRL,Dincao10}
but were restricted to small confinements or assumed no transitions between confining states.
In the present formalism, we have no such restrictions.
Our method is based on a frame transformation between spherical to cylindrical 
coordinates, similar to that employed in Ref.~\cite{Granger04, Omahony91} for example.
The frame transformation has the advantage of treating in full detail the microscopic physics of the molecule-molecule
interaction, while projecting onto appropriate two dimensional scattering states.
We consider two ultracold polar molecules of masses $m_1,m_2$ and positions $\vec{r}_1, \vec{r}_2$  
from a fixed arbitrary origin O (see Fig.~\ref{COORD-FIG}-a).
The molecules are confined in a harmonic oscillator trap $V_{\text{ho}}^{\tau}  = m_\tau \, \omega^2 \, z_\tau^2/2$ 
for molecule $\tau=1,2$, 
of angular frequency $\omega = 2 \pi \nu$.
An electric field applied along the confinement direction $\hat{z}$
polarizes the molecules, giving them dipole moments $\vec{d}_\tau = d_\tau \, \hat{z}$.
We use cartesian coordinates $(x_\tau, y_\tau, z_\tau)$ to describe the vector $\vec{r}_\tau$.
We also use
the center-of-mass (CM) coordinate $\vec{R} = (m_1 \vec{r}_1 + m_2 \vec{r}_2) / (m_1+m_2)$
and the relative coordinate $\vec{r} = \vec{r}_2 - \vec{r}_1$ (see Fig.~\ref{COORD-FIG}-a).
We use the cartesian coordinate $(X, Y, Z)$ to describe the vector $\vec{R}$,
and either cylindrical coordinates $(\rho, z, \varphi)$ 
or spherical coordinates $(r, \theta, \varphi)$
to describe the vector $\vec{r}$ (see Fig.~\ref{COORD-FIG}-b),
with $\rho = r \, \sin \theta$ and $z = r \, \cos \theta$.
Both the electric field and the harmonic oscillator potential are applied along the $z$ axis, which we take as
the quantization axis.

\subsection{Hamiltonian}

The total Hamiltonian of the system is
\begin{eqnarray}
H_\text{tot} &=&  {T}_1 + {T}_2  + {V}
\label{Hamiltoniantot}
\end{eqnarray}
with ${T}_\tau = - \hbar^2 \nabla^2_{\vec{r}_\tau} /(2 m_\tau)$ representing the kinetic energy operator
of the molecule $\tau$. $V$, the potential energy, is given by
%
%
\begin{eqnarray}
{V}  &=&  V_{\text{abs}} + V_{\text{vdW}} +  V_{\text{dd}} +  V_{\text{ho}}^{\tau=1} + V_{\text{ho}}^{\tau=2}  \nonumber \\
   &=& i A e^{-(r-r_\text{min})/r_\text{c}} - \frac{C_6}{r^6} +  \frac{d_1 \, d_2 \, (1 - 3 \cos^2{\theta})}{4 \pi \varepsilon_0 \, r^3} \nonumber \\
 &+&  \frac{1}{2} \bigg( m_1 \, \omega^2 \, z_1^2 + m_2 \, \omega^2 \, z_2^2 \bigg).
\label{potential}
\end{eqnarray}
%
%
The first term on the right hand side represents
an appropriate imaginary potential capturing the overall chemical couplings at short-range.
It replaces ab initio calculations of the electronic structure
of trimer and tetramer alkali complexes, which remain incomplete for KRb~\cite{Zuchowski10,Byrd10,Soldan10,Meyer10}.
For the time being, an absorbing potential
has shown very good agreement
with experimental results~\cite{Ospelkaus10-SCIENCE,Ni10-NATURE,Idziaszek10-PRL,Idziaszek10-PRA}
for KRb molecules. We use the same absorbing potential here.
The second term represents
an isotropic van der Waals interaction ; the third term represents the dipole-dipole interaction
where $d_\tau$ represents the expectation value in the $z$ direction of the dipole moment
induced by the electric field ; 
and
the last two terms represent the one dimensional harmonic oscillator trap
that confines the molecules in a plane perpendicular to the $z$ direction.
The initial energy of the molecule $\tau$ in the trap is given
by $\varepsilon_{n_\tau}=\hbar \omega (n_\tau+1/2)$,
where $n_\tau$ represents the associated quantum number
of the harmonic oscillator state they are loaded into.
The associated function
is the usual normalized eigenfunction of the harmonic oscillator $g_{n_\tau}(z_\tau)$.

\subsection{Symmetrized internal and external states}

We consider here identical molecules with  
same masses ($m_1=m_2$) and same dipoles ($d_1=d_2=d$).
As the molecules are identical, we have to construct
an overall wavefunction $\Psi$ of the system
for which the molecular permutation operator $P$
gives  
\begin{eqnarray}
P \, \Psi = \epsilon_P \, \Psi
\label{sym}
\end{eqnarray}
with $\epsilon_P=+1$
for bosonic molecules and $\epsilon_P=-1$
for fermionic molecules.
This overall wavefunction $\Psi$ is constructed from an internal wavefunction $|\alpha_1 \, \alpha_2 \rangle$ representing 
the electronic, vibrational, rotational and spin degrees of freedom of molecule 1 and 2 respectivelly ;
from an external wavefunction $|n_1 \, n_2 \rangle$ representing 
the one dimensional individual confining wavefunction $g_{n_1}(z_1) \, g_{n_2}(z_2)$  ;
and finally from a two dimensional collision wavefunction in the plane perpendicular to the confinement.

We first build symmetrized states of the internal wavefunction
\begin{eqnarray}
|\alpha_1 \, \alpha_2, \eta \rangle = \frac{1}{\sqrt{2(1+\delta_{\alpha_1,\alpha_2})}} \bigg[ |\alpha_1 \, \alpha_2 \rangle + \eta |\alpha_2 \, \alpha_1 \rangle \bigg]
\end{eqnarray}
for which $P \, |\alpha_1 \, \alpha_2, \eta \rangle = \eta \, |\alpha_1 \, \alpha_2, \eta \rangle$.
$\eta$ is a good quantum number and is conserved during the collision.
If the molecules are in the same molecular internal state, 
only the symmetry $\eta=+1$ has to be considered. If they are in different internal state,
both symmetries $\eta=\pm1$ have to be considered. 
We omit explicit reference to the internal wavefunctions $|\alpha_1 \, \alpha_2, \eta \rangle $ in the following,
but the quantum number $\eta$ still plays a role in the selection rules, as discussed in Appendix C.

We next build symmetrized states of the external confining wavefunction
\begin{eqnarray}
|n_1 \, n_2, \gamma \rangle = \frac{1}{\sqrt{2(1+\delta_{n_1,n_2})}} \bigg[ |n_1 \, n_2 \rangle + \gamma |n_2 \, n_1 \rangle \bigg]
\label{symstaten1n2}
\end{eqnarray}
with $P \, |n_1 \, n_2, \gamma \rangle = \gamma \, |n_1 \, n_2, \gamma \rangle $.
$\gamma$ is a good quantum number and is conserved during the collision.
If the molecules are in the same external confining state,
only the symmetry $\gamma=+1$ has to be considered. If they are in different external state,
both symmetries $\gamma=\pm1$ have to be considered.
It is useful at this point to turn into 
a relative/CM representation of the confining states.
It is easy to show that
the Hamiltonian~\eqref{Hamiltoniantot} 
can also be written in the relative/CM representation as
%
%
\begin{multline}
H_\text{tot} =  {T}_{\text{rel}} + {T}_{\text{CM}}  + V_{\text{abs}}  
+ V_{\text{vdW}}  \\
+  V_{\text{dd}} +  V^\text{rel}_\text{ho}  + V^\text{CM}_\text{ho}
\label{Hamiltoniantotrel}
\end{multline}
with ${T}_{\text{rel}} = - \hbar^2 \nabla^2_{\vec{r}} /(2 \mu)$ 
and ${T}_{\text{CM}} = - \hbar^2 \nabla^2_{\vec{R}} /(2 m_{\text{tot}})$,
$\mu = m_1 m_2 / (m_1 + m_2)$ and $m_\text{tot} = m_1 + m_2$, 
$V^\text{rel}_\text{ho} = \mu \, \omega^2 \, z^2/2$ and
$V^\text{CM}_\text{ho} = m_\text{tot} \, \omega^2 \, Z^2/2$.
The associated energies and functions will be denoted $\varepsilon_{n}, \varepsilon_{N}$
and $g_{n}(z),g_{N}(Z)$.
These harmonic oscillator states in the relative and CM coordinates are related to 
those in independent particle coordinates $g_{n_1}(z_1), g_{n_2}(z_2)$ by
(see Appendix A)
%
%
\begin{multline}
g_{n_1}(z_1) \, g_{n_2}(z_2) = \frac{1}{\sqrt{2^{2(n_1+n_2)} \, n_1! \, n_2!}} \\
\sum_{k=0}^{n_1}  \sum_{k'=0}^{n_2}  \sum_{q=0}^{\text{min}(k,k')}  \sum_{q'=0}^{\text{min}(n_1-k,n_2-k')} \\
\frac{n_1! \, n_2!}{(k-q)! \, (k'-q)! \, q! \, (n_1-k-q')! \, q'! \, (n_2-k'-q')!}  \\
\times (-1)^{n_1-k} \, 2^q \, 2^{q'} \, \sqrt{2^n n!} \, \sqrt{2^N N!} \, g_{n}(z) \, g_{N}(Z) 
\label{gNgntogn1gn2}
\end{multline}
with
\begin{eqnarray}
n &=& -2q' + n_1 + n_2 - k - k' \nonumber \\
N &=& -2q + k + k'.
\end{eqnarray}
We give in Appendix A explicit relations between $|n_1 \, n_2 \rangle$ 
and $|n, \, N \rangle$ states for low values of quantum numbers, $0 \le n_1, n_2 \le 2$,
and in Appendix B the relations between the symmetrized individual representation $|n_1 \, n_2, \gamma \rangle$ 
and the relative/CM representation $|n, \, N \rangle$ states, using Eq.~\eqref{gNgntogn1gn2} and Eq.~\eqref{symstaten1n2}.

\subsection{Diabatic-by-sector method}

To solve the Schr\"odinger equation for $\Psi$, we work in the relative/CM representation $|n , N \rangle$ since we know
how to come back to the physical $|n_1 \, n_2, \gamma \rangle$ representation.
In the relative/CM representation,
the collisional problem depends only on the coordinate $Z$ and the relative vector $\vec{r}$, 
and not on the coordinates $X$ and $Y$.
In the following, we explicitly remove these two coordinates from the problem.
If we use the coordinate $Z$ and spherical coordinates to represent $\vec{r}$, 
the Hamiltonian is given by 
%
%
\begin{multline}
H  =  - \frac{\hbar^2}{2 \mu} \frac{1}{r^2} \, \frac{\partial}{\partial r} \, \bigg( r^2 \, \frac{\partial}{\partial r} \bigg) 
+ \frac{\hat{L}^2}{2 \mu r^2} 
+ V_{\text{abs}} + V_\text{vdW} + V_\text{dd} \\
+ V^\text{rel}_\text{ho} 
- \frac{\hbar^2}{2 m_\text{tot}} \frac{\partial^2}{\partial Z^2} + V^\text{CM}_\text{ho} .
\label{Hamiltonianrelsph}
\end{multline}
If we use the coordinate $Z$ and cylindrical coordinates to represent $\vec{r}$,           
the Hamiltonian is given by 
\begin{multline}
H  = - \frac{\hbar^2}{2 \mu} \, \left\{ \frac{\partial^2}{\partial \rho^2}  + \frac{1}{\rho} \, \frac{\partial}{\partial \rho} 
+  \frac{1}{\rho^2} \, \frac{\partial^2}{\partial \varphi^2} \right \} 
+ V_{\text{abs}} + V_\text{vdW} + V_\text{dd}  \\
- \frac{\hbar^2}{2 \mu} \, \frac{\partial^2}{\partial z^2}  + V^\text{rel}_\text{ho} 
- \frac{\hbar^2}{2 m_\text{tot}} \frac{\partial^2}{\partial Z^2} + V^\text{CM}_\text{ho} .
\label{Hamiltonianrelcyl}
\end{multline}

In a diabatic-by-sector method~\cite{Pack87,Launay89,Quemener06-PHD},
using a spherical coordinate representation of the wavefunction,
the range over the Schr\"odinger equation to be solved, $r_{\text{min}} \le r \le r_{\text{max}}$,
is divided into $N_s$ sectors of width $\Delta r = (r_{\text{max}} - r_{\text{min}}) / N_s$.
The middle of each sector corresponds to a grid point $r_p$, with $p=1,...,N_s$.
At each grid point $r=r_p$, we use $N_l$ normalized Legendre 
polynomials $P^{M_L}_{L}(\cos \theta)$ for a given value of $M_L$, the 
quantum number associated with the azimuthal projection
of the orbital angular momentum $\hat{L}$ on the $z$ direction,
to diagonalize
the angular Hamiltonian ${\cal H}^{M_L,\eta}(r, \theta) = \hat{L}^2/(2 \mu r^2)
+ V_{\text{abs}} + V_\text{vdW} + V_\text{dd} + V^{\text{rel}}_{\text{ho}}$ 
of the Hamiltonian in Eq.~\eqref{Hamiltonianrelsph}.
The resulting eigenfunctions are the adiabatic functions $\chi^{M_L,\eta}_{j}(r_p;\theta)$ with $j=1,...,N_l$. 
They are used as a basis set for the representation of the total wavefunction
%
%
\begin{multline}
\Psi^{M_L,\eta,N}_{j}(r,\theta,\varphi,Z) =   \\ 
\frac{1}{r} \, \sum_{j''=1}^{N_\text{adiab}}  \chi^{M_L,\eta}_{j''}(r_p;\theta) \, g_{N}(Z) \,  F^{M_L,\eta,N}_{j'' j}(r_p;r) \, \frac{e^{iM_L\varphi}}{\sqrt{2\pi}}
\label{Psisph}
\end{multline}
for a given adiabatic state $j$.
The associated eigenenergies of the angular hamiltonian are the adiabatic energies $\varepsilon_j(r_p)$.
They converge to the relative harmonic oscillator
energies $\varepsilon_{n}$ with $n=0,...,N_l-1$ at large $r_p$,
so that a one-to-one correspondence can be identified between the adiabatic quantum states $j=1,...,N_l$
and the relative harmonic oscillator quantum states $n=0,...,N_l-1$.
In practice, we use a truncated number of adiabatic functions $N_\text{adiab} \ll N_l$.
If we restrict the independent oscillator quantum numbers $0 \le n_1,n_2 < n_{\text{osc}}^\text{max}$,
then the maximum value that the relative quantum number $n$ can take is $2 \, n_{\text{osc}}^\text{max}$
and we choose
$N_\text{adiab} = 2 \, n_{\text{osc}}^\text{max}$.
In Eq.~\eqref{Psisph}, we use the fact that
there are no terms in~\eqref{Hamiltonianrelsph}
that create mixings between different values of $N$.
Moreover, the potential $V$
does not depend on the azimuthal angle $\varphi$.
As a consequence, the quantum numbers $N$ and  $M_L$
are conserved during the collision. 

The total energy $E$ is equal to $\varepsilon_{n_1} + \varepsilon_{n_2} + E_c$, 
where $\varepsilon_{n_1}, \varepsilon_{n_2}$
are the energies of the molecules $1,2$ in the confining potential,
when they start initialy in $n_1, n_2$,
and $E_c$ is the initial collision energy between the two molecules
in the two dimensional plane. 
$E$ is conserved during the collision.
Solving the time-independent Schr\"odinger equation $H \Psi = E \Psi$ 
provides the following set of close-coupling differential equations in spherical coordinates
for each values of $M_L$, $\eta$ and $N$,
from a state $j$ to a state $j'$
%
%
\begin{multline}
\left\{ - \frac{\hbar^2}{2 \mu} \frac{d^2}{d r^2}
 + \varepsilon_{N} - E \right\}
\, F^{M_L,\eta,N}_{j' j}(r_p;r)  \\ 
+ \sum_{j''=1}^{N_\text{adiab}}
  {\cal U}^{M_L,\eta}_{j' j''}(r_p;r)
\  F^{M_L,\eta,N}_{j''  j}(r_p;r) = 0
\label{eqcoup}
\end{multline}
where
%
%
\begin{multline}
{\cal U}^{M_L,\eta}_{j' j''}(r_p;r)
 =   \\
\int_0^{\pi} \,
\chi^{M_L,\eta}_{j'}(r_p;\theta)
\, {\cal H}^{M_L,\eta}(r, \theta)
\, \chi^{M_L,\eta}_{j''}(r_p;\theta)  \, \sin \theta \, d\theta .
\label{Umatrix}
\end{multline}
The goal is to find all the elements $F^{M_L,\eta,N}_{j' j}$.
We employ the standard method of the propagation of the log-derivative
matrix~\cite{Johnson73}
%
%
\begin{multline}
Z^{M_L,\eta,N}(r_p;r) =  \\
\bigg\{ (\partial / \partial r)  F^{M_L,\eta,N}(r_p;r) \bigg\} \bigg\{ F^{M_L,\eta,N}(r_p;r) \bigg\}^{-1}
\end{multline}
with matrix elements $Z^{M_L,\eta,N}_{j'  j}(r_p;r)$,
and obtain these elements for all possible 
states $j$ to all possible states $j'$. 
In the diabatic-by-sector method, one has to perform a transformation 
operation from sectors to sectors, since the adiabatic
functions $\chi^{M_L,\eta}(r_p;\theta)$ change from $r_p$ to $r_{p+1}$.
Then the log-derivative expressed in the basis of the sector $p+1$
at the distance $ r = r_p + \Delta r / 2$ separating the sector $p$ and $p+1$
is given by
%
%
\begin{multline}
Z^{M_L,\eta,N}(r_{p+1};r=r_p + \Delta r / 2) 
=  \\
P \ Z^{M_L,\eta,N}(r_{p};r=r_p + \Delta r / 2) \ P^{-1}
\end{multline}
with the passage matrix
\begin{eqnarray}
P_{j' j} =  \int_0^{\pi} \,
\chi^{M_L,\eta}_{j'}(r_{p+1};\theta)
\, \chi^{M_L,\eta}_{j}(r_p;\theta)  \, \sin \theta \, d\theta .
\end{eqnarray}

\subsection{Asymptotic matching}

Compared to free molecules in 3D,
the external confinement  $V^\text{rel}_\text{ho}$ in Eq.~\eqref{Hamiltonianrelsph}
persists
at large intermolecular separation $r$,
and the spherical representation of $\vec{r}$ is not appropriate anymore.
Instead, we use in the asymptotic region cylindrical coordinates 
appropriate to the potential $V^\text{rel}_\text{ho}$.
For a given state of relative quantum number $n$,
we now expand the total wavefunction as follows
%
%
\begin{multline}
\Psi^{M_L,\eta,\gamma,N}_{n}(\rho,z,\varphi,Z) =  \\
  \frac{1}{\rho^{1/2}} \, \sum_{n''} g_{n''}(z)  \, g_{N}(Z) \, G^{M_L,\eta,\gamma,N}_{n'' n}(\rho) \,  \frac{e^{iM_L\varphi}}{\sqrt{2\pi}}.
\label{Psicyl}
\end{multline}
In the following, we will use the short-hand notation $\xi \equiv M_L,\eta,\gamma,N$.
Note that because we use the coordinate $Z$ and the wavefunction $g_{N}(Z)$ in both spherical and cylindrical
representation, the external confinement $V^\text{CM}_\text{ho}$ 
is always well described.
When $\rho \to \infty$, $ V_{\text{abs}} + V_\text{vdW} + V_\text{dd} \to 0$, and
the close-coupling asymptotic Schr\"odinger equations become
%
%
\begin{multline}
\bigg\{ - \frac{\hbar^2}{2 \mu} \frac{d^2}{d \rho^2} +
   \frac{\hbar^2 \, (M_L^2-1/4)}{2 \mu \rho^2} \\
+ \varepsilon_{n} + \varepsilon_N - E \bigg\}
\, G_{n' n}^{\xi}(\rho)
= 0.
\label{asymSchrodinger2D}
\end{multline}
At large $\rho$, the radial function $G_{n'  n}^{\xi}(\rho)$ in Eq.~\eqref{Psicyl}
is a linear combination of two possible solutions 
$G_{n'}^{\xi(1,2)}(\rho)$ of Eq.~\eqref{asymSchrodinger2D},
and takes the form
\begin{eqnarray} 
G_{n'  n}^{\xi}(\rho)  {\raisebox{-2.mm}{$\;\stackrel{{\textstyle\longrightarrow}}
{{\scriptstyle {\rho} \to \infty}}\;$}} 
G_{n'}^{\xi,(1)}(\rho) \, \delta_{n,n'} 
+ G_{n'}^{\xi,(2)}(\rho) \, K^{\xi}_{n'  n} .
\label{condlimK}
\end{eqnarray}
$K^{\xi}_{n'  n}$ represents
an element of the reactance matrix.
The functions $G_{n'}^{\xi,(1,2)}$
represent the regular and irregular asympotic solutions of the radial Schr\"odinger equation Eq.~\eqref{asymSchrodinger2D}
\begin{eqnarray} 
G_{n'}^{\xi,(1)}(\rho) &=& \rho^{1/2} \, J_{M_L}(k_{n',N} \, \rho) \nonumber \\
G_{n'}^{\xi,(2)}(\rho) &=& \rho^{1/2} \, N_{M_L}(k_{n',N} \, \rho) 
\label{Bessel}
\end{eqnarray}
where $J_{M_L}, N_{M_L}$ are Bessel functions~\cite{AS}
and $k_{n',N}=\sqrt{2 \, \mu \, (E - \varepsilon_{n'} - \varepsilon_{N})}/\hbar$
is the wave number in the channel $n'$ of the relative harmonic oscillator.
If $E - \varepsilon_{n'} - \varepsilon_{N} < 0$,
the modified Bessel functions have to be used instead.

To determine $K$, we must transform between the spherical wavefunction
that captures the short-range physics and the cylindrical wavefunction
that captures the asymptotic boundary conditions.
The regular and irregular spherical radial functions $F^{\xi,(1,2)}(r_p;r)$
and their derivatives
can be connected to their cylindrical asymptotic counterpart $G^{\xi,(1,2)}(\rho)$ by
equating the wavefunctions Eq.~\eqref{Psisph} and Eq.~\eqref{Psicyl} and their derivatives at
a constant sphere of radius $r=r_\text{max}$
%
%
\begin{multline}
F_{j' j}^{\xi,(1,2)}(r_{p=N_s};r) \  \bigg|_{r = r_\text{max}}
 =  \int_0^{\pi} \,
\chi^{M_L,\eta}_{j'}(r_{p=N_s};\theta) \\
\, \frac{r}{\rho^{1/2}} \, g_{n}(z) \, G_{n}^{\xi,(1,2)}(\rho)
\, \sin\theta \, d\theta  \ \bigg|_{r = r_\text{max}}   
\label{matching2D-1}
\end{multline}
\begin{multline}
\frac{\partial}{\partial r} \bigg( F_{j' j}^{\xi,(1,2)}(r_{p=N_s};r) \bigg) \ \bigg|_{r = r_\text{max}}
 =  \int_0^{\pi} \,
\chi^{M_L,\eta}_{j'}(r_{p=N_s};\theta) \\
\, \frac{\partial}{\partial r} \bigg\{  \frac{r}{\rho^{1/2}} \, g_{n}(z) \, G_{n}^{\xi,(1,2)}(\rho) \bigg\}
\, \sin\theta \, d\theta   \ \bigg|_{r = r_\text{max}} 
\label{matching2D-2}
\end{multline}
with the one-to-one correspondence 
$\{n = 0, ... , N_\text{adiab}-1\} \equiv \{j = 1, ..., N_\text{adiab}\}$
between the quantum numbers $n$ and $j$. $r_{p=N_s}$ is the middle of the last sector $N_s$.
This is a similar matching procedure
that connects short-range democratic hyperspherical coordinates
to asymptotic Jacobi coordinates
employed
in atom-molecule chemical reactive scattering studies~\cite{Pack87,Launay89,Quemener06-PHD}.
Convergence with respect to $N_\text{adiab}$ and $r_\text{max}$ is found when the Wronskian matrix 
\begin{eqnarray}
F^{\xi,(1)} \frac{\partial}{\partial r} \bigg( F^{\xi,(2)} \bigg)
- \frac{\partial}{\partial r} \bigg( F^{\xi,(1)} \bigg) F^{\xi,(2)} 
\label{wronskian}
\end{eqnarray}
converges to the unit matrix.

The $K$ matrix is determined by the matrix operation
\begin{eqnarray} 
K^{\xi} = - \frac{Z^{\xi} F^{\xi,(1)}-(\partial / \partial r)(F^{\xi,(1)})}{Z^{\xi} F^{\xi,(2)}-(\partial / \partial r)(F^{\xi,(2)})}. 
\label{Kmat}
\end{eqnarray}
The scattering matrix ${\cal S}$ in the relative/CM representation is determined by
\begin{eqnarray} 
{\cal S}^{\xi} = \frac{I - iK^{\xi}}{I + iK^{\xi}} 
\label{Smat}
\end{eqnarray}
where in this equation, $I$ represents the unit matrix.
The scattering matrix in the symmetrized individual representation 
$|n_1 \, n_2 , \gamma \rangle$ is found by gathering all individual scattering matrices ${\cal S}$ 
corresponding to different values of $N$ and by applying a transformation from the relative/CM representation
to the symmetrized individual representation 
\begin{eqnarray} 
S^{M_L,\eta,\gamma} = U \, \bigg\{ \sum_{N}^{\bigoplus} {\cal S}^{M_L,\eta,\gamma,N} \bigg\} \, U^{T}.
\label{Smattransform}
\end{eqnarray}
The transformation matrix $U$, with elements $ U_{n_1 \, n_2 , \gamma  ;  n, N} = \langle n_1 \, n_2 , \gamma  | n, N  \rangle$,
can be found using the relations in Appendix B.
We use the transpose $U^{T}$ of the matrix $U$ instead of its inverse because $U$
is not generally a square matrix.

\subsection{Observables}

After a collision, the quantum probability from an initial 
state $n_{1} \, n_{2} $ to a final state 
$n_{1}' \, n_{2}' $ 
for defined numbers $M_L, \eta, \gamma$
is given by $P^{M_L,\eta,\gamma}_{n_1' \, n_2', n_1 \, n_2} = |S^{M_L,\eta,\gamma}_{n_1' \, n_2', n_1 \, n_2}|^2$.
The elastic, inelastic (confining state changing) and reactive probabilities are given by
\begin{eqnarray} 
P^{\text{el}, \, M_L,\eta,\gamma} &=& P^{M_L,\eta,\gamma}_{n_1 \, n_2, n_1 \, n_2} \nonumber \\
P^{\text{in}, \, M_L,\eta,\gamma} &=& \sum_{n_1' \, n_2' \ne n_1 \, n_2} P^{M_L,\eta,\gamma}_{n_1' \, n_2', n_1 \, n_2} \nonumber \\
P^{\text{re}, \, M_L,\eta,\gamma} &=& 1 - P^{\text{el}, \, M_L,\eta,\gamma} - P^{\text{in},\, M_L,\eta,\gamma}.
\label{proba}
\end{eqnarray}
We mean by ``inelastic'', processes that change the external confining states of the molecules. 
Finally, for an initial state $n_{1} \, n_{2} $,
the elastic, inelastic, 
and reactive cross sections are given by~\cite{Lapidus82,Adhikari86,Naidon06}
\begin{eqnarray}
\sigma_{n_1 \, n_2}^\text{el} &=&  \frac{\hbar}{\sqrt{2 \mu E_c}} \sum_{M_L,\eta,\gamma} |1 - S^{M_L,\eta,\gamma}_{n_1 \, n_2, n_1 \, n_2}|^2 \times \Delta   \nonumber \\ 
\sigma_{n_1 \, n_2}^\text{in} &=&  \frac{\hbar}{\sqrt{2 \mu E_c}} \sum_{M_L,\eta,\gamma} P^{\text{in}, \, M_L,\eta,\gamma} \times \Delta   \nonumber \\ 
\sigma_{n_1 \, n_2}^\text{re} &=&  \frac{\hbar}{\sqrt{2 \mu E_c}} \sum_{M_L,\eta,\gamma} P^{\text{re}, \, M_L,\eta,\gamma} \times \Delta .
\label{cross}
\end{eqnarray}
The inelastic state-to-state cross section is given by
\begin{eqnarray}
\sigma_{n_1 \, n_2 \, \text{to} \, n_1' \, n_2'}^\text{in}
&=&  \frac{\hbar}{\sqrt{2 \mu E_c}} \sum_{M_L,\eta,\gamma} P^{M_L,\eta,\gamma}_{n_1' \, n_2', n_1 \, n_2}  \times \Delta . 
\label{crossst2st}
\end{eqnarray}
The factor $\Delta$ 
represents symmetrization requirements for indistinguishable particles
in a same internal and confining state~\cite{Burke99,Quemener10-QT}.
The cross sections
are found by summing over all the contributions of different values of $M_L,\eta,\gamma$.
For the ultralow energies involved in this study, only
the first partial wave will be required for indistinguishable molecules
(same internal states $\eta=+1$ and same confining state $\gamma=+1$): 
the $M_L=0$ partial wave for indistinguishable bosons
and the $M_L=\pm1$ partial wave for indistinguishable fermions.
The temperature dependence of the loss rates in the two dimensional plane is found by averaging the cross sections over a 
two-dimensional Maxwell--Boltzmann distribution of the relative velocity $v = \sqrt{2 E_c / \mu}$
in the two-dimensional plane.
This gives a two dimensional thermalized rate
\begin{eqnarray}
\beta_{n_1 \, n_2}^{T,\text{el,in,re}} =  \int_0^\infty \sigma_{n_1 \, n_2}^\text{el,in,re} \, v \, f(v) \, dv 
\label{rate2DT}
\end{eqnarray}
with
\begin{eqnarray}
f(v) =  \frac{\mu}{k_B T} \, v \, e^{-\frac{\mu v^2}{2 k_B T}}
\label{distMB2D}
\end{eqnarray}
where $k_B$ is the Boltzmann constant.
The rate in Eq.~\ref{rate2DT} corresponds to the rate per molecule, 
not the event or collision rate~\cite{Burke99,Quemener10-QT}.

Selection rules apply due to symmetrization of the wavefunction
under permutation of identical molecules (Appendix C).
The rules are
\begin{eqnarray}
\eta \, \gamma \, (-1)^{M_L} = \eta \, (-1)^L = \eta \, (-1)^{M_L+n} = \epsilon_P .
\label{selectionrule}
\end{eqnarray}
This limits the summation over $M_L,\eta,\gamma$ in Eq.~\eqref{cross} and~\eqref{rate2DT}
and the values of the quantum numbers $j''$ and $n''$ used in Eq.~\eqref{Psisph} and Eq.~\eqref{Psicyl}.\\

In the following, we will consider molecules of KRb as an illustrative example
of experimental interest~\cite{Ospelkaus10-SCIENCE,Ni10-NATURE,Miranda10,Aikawa10-ARXIV}.
For concreteness, we will take the isotope $^{39}$K$^{87}$Rb for the bosonic molecules ;
the results for the bosonic isotope $^{41}$K$^{87}$Rb~\cite{Aikawa10-ARXIV}
are nearly identical.
We take the isotope $^{40}$K$^{87}$Rb for the fermionic molecules~\cite{Ospelkaus10-SCIENCE,Ni10-NATURE,Miranda10}.
Convergence of the results have been checked with the matching distance $r_\text{max}$
and the number of adiabatic functions $N_\text{adiab}$ included in the expansion of the wavefunction.
Unless stated otherwise, 
we choose $r_\text{min}=10 \ a_0$  and $r_\text{max}=10000 \ a_0$ ($a_0 \simeq 0.529$~Angstroms is the Bohr radius), 
$N_s=10000$ sectors, $0 < n_1, n_2 < n_\text{osc}^\text{max}=3$, 
$N_\text{adiab} = 2 \, n_\text{osc}^\text{max} = 6$ and only the first partial waves $M_L=0,1$
depending on the species and the selection rules involved.
We used $N_l = 80$ Legendre polynomials for $\nu < 100$~kHz and $N_l = 120$ for $\nu \ge 100$~kHz,
to construct the adiabatic functions.
This yields converged results of 10~\% at most for the elastic rates (more especially at high confinement)
and 1~\% for the reactive and inelastic rates.
For $V_\text{abs}$, we use $A=-10$~K and $r_c=10 \ a_0$, which adequately reproduces 
experimental loss rates in three dimensional collisions~\cite{Ni10-NATURE}.

\section{Suppression of chemical reactions}

\begin{figure} [t]
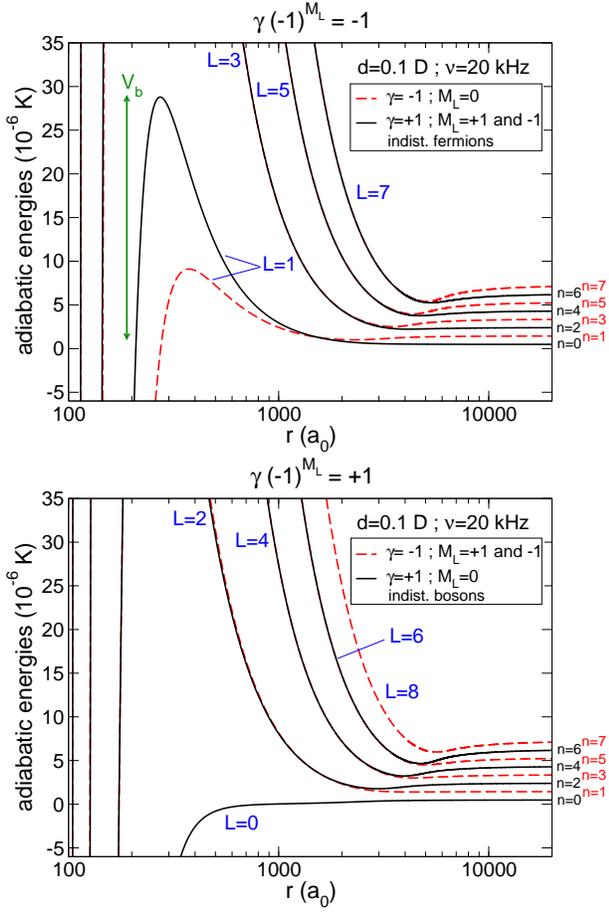

\begin{center}
\includegraphics*[width=8cm,keepaspectratio=true,angle=0]{figure2a.eps}  \\
\includegraphics*[width=8cm,keepaspectratio=true,angle=0]{figure2b.eps}
\caption{(Color online) Adiabatic energies versus $r$ for the $\gamma \, (-1)^{M_L} = -1$ symmetry (upper panel) and
for the $\gamma \, (-1)^{M_L} = +1$ symmetry (lower panel), for $\nu = 20$~kHz and $d=0.1$~D.
The black (red dashed) curves correspond to $\gamma = +1$ ($\gamma = - 1$) manifolds.
We also show how values of $L$ and $n$ adiabatically connect.
$V_b$ is the height of the barrier for molecules in the lowest confining state ($n=0$).
\label{SPAG-FIG}
}
\end{center}
\end{figure}

We discuss in this section how chemical reactions proceed when the reactants are 
subject to different confinements and electric fields.
We present in Fig.~\ref{SPAG-FIG} the adiabatic energies $\varepsilon_j(r_p)$ 
for the symmetry $\gamma \, (-1)^{M_L} = -1$ (upper panel) and the symmetry $\gamma \, (-1)^{M_L} = +1$ (lower panel), 
for a trap with $\nu = 20$~kHz and induced dipole moment $d=0.1$~D.
These energies converge at large $r$ to the energies of the relative harmonic oscillator $\varepsilon_{n}$.
To associate a specific confined collision with a symmetry $\gamma \, (-1)^{M_L} $, one has to use
Eq.~\eqref{selectionrule}. 
If the molecules are identical fermions in the same internal state, 
$\eta = +1$ and $\epsilon_P=-1$, and then $\gamma \, (-1)^{M_L} = -1$,
so the scattering problem only employs the black and red dashed curves 
of the upper panel in Fig.~\ref{SPAG-FIG}.
In addition, if the identical fermionic molecules are in the same external state, then $\gamma=+1$,
and the scattering problem only uses the black curves.
If however the identical fermionic molecules are in different internal states,
both values of $\eta$ are relevant.
Then, in the case of $\eta=-1$, now $\gamma \, (-1)^{M_L} = +1$,
and the black and red dashed curves of the lower panel have to be employed as well.
If the fermionic molecules are in different internal states, but in the same external state,
then $\gamma=+1$, and
one has to use only the black curves of both panels.

Using similar arguments, if molecules are identical bosons in the same internal state, 
one has to use the black and red dashed curves of the lower panel. 
If besides they are in the same external state,
only the black curves have to be used.
If they are in different internal states, 
all black and red dashed curves of both panels have to be used, while only the black curves
of both panels are used if the identical bosons are in different internal states but in the same external state.
The case of two different polar molecules corresponds to all curves of all symmetries employed.
Also, note that because $\gamma \, (-1)^{M_L} = (-1)^L = (-1)^{M_L+n}$ in Eq.~\eqref{selectionrule},
the values of $L$ and $M_L+n$ are odd for the upper panel and even for the lower panel,
and the $\gamma=+1$ ($\gamma=-1$) curves corresponds to even (odd) relative quantum numbers $n$ ($\gamma = (-1)^{n}$).
Therefore, symmetry consideration are essential for the dynamics of ultracold molecules
in confined geometry and electric field. \\

\begin{figure} [t]
\begin{center}
\includegraphics*[bb=100 0 650 480,width=9cm,keepaspectratio=true,angle=0]{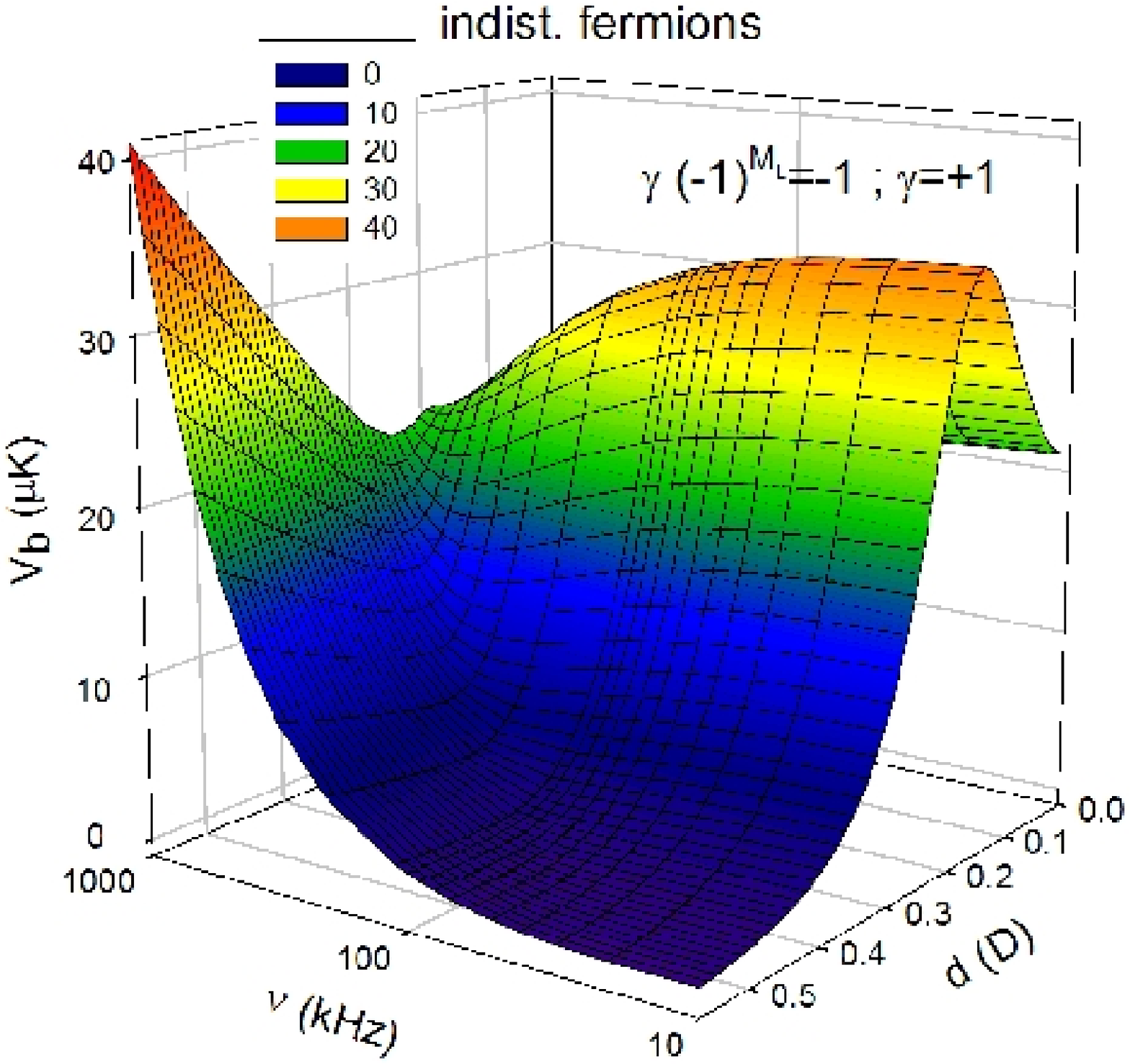} \\
\includegraphics*[bb=100 0 650 480,width=9cm,keepaspectratio=true,angle=0]{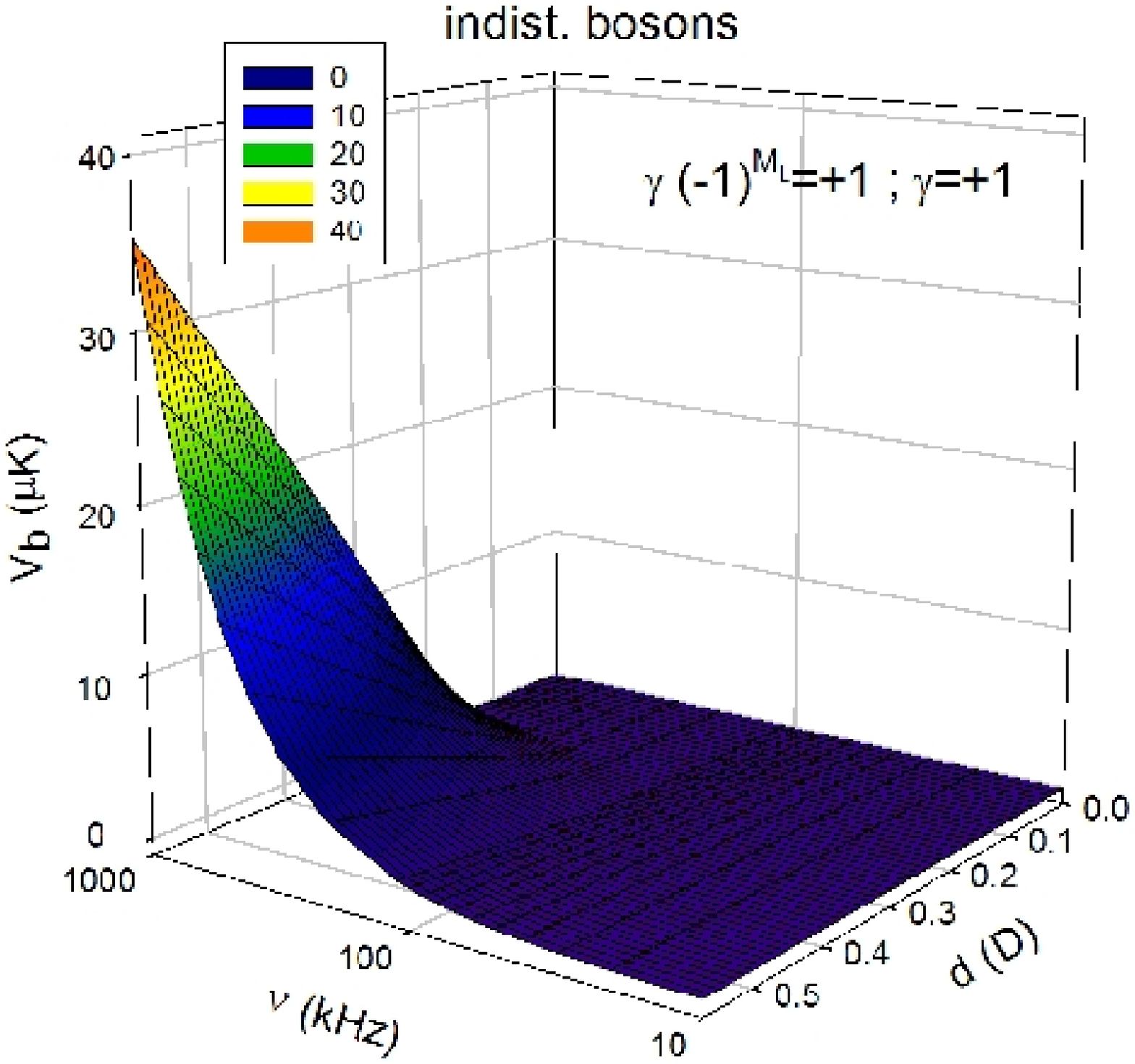}
\caption{(Color online) Height of the adiabatic barrier $V_b$ versus $d$ and $\nu$ for indistinguishable fermions (upper panel) and
for indistinguishable bosons (lower panel) in the lowest confining state.
\label{BARRIER-FIG}
}
\end{center}
\end{figure}

\begin{figure} [t]
\begin{center}
\includegraphics*[bb=20 0 410 380,width=8cm,keepaspectratio=true,angle=0]{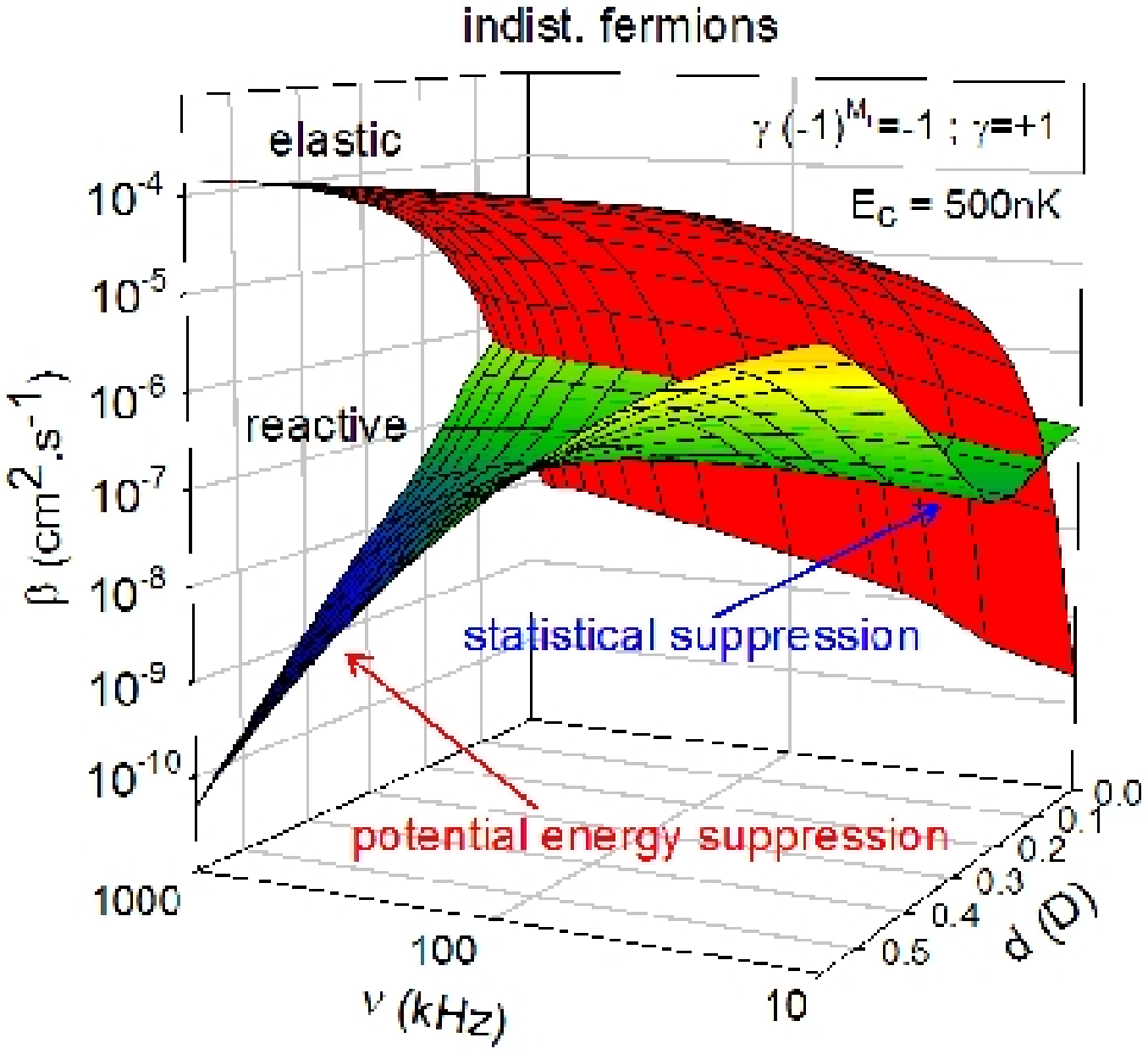} \\
\includegraphics*[bb=20 0 410 380,width=8cm,keepaspectratio=true,angle=0]{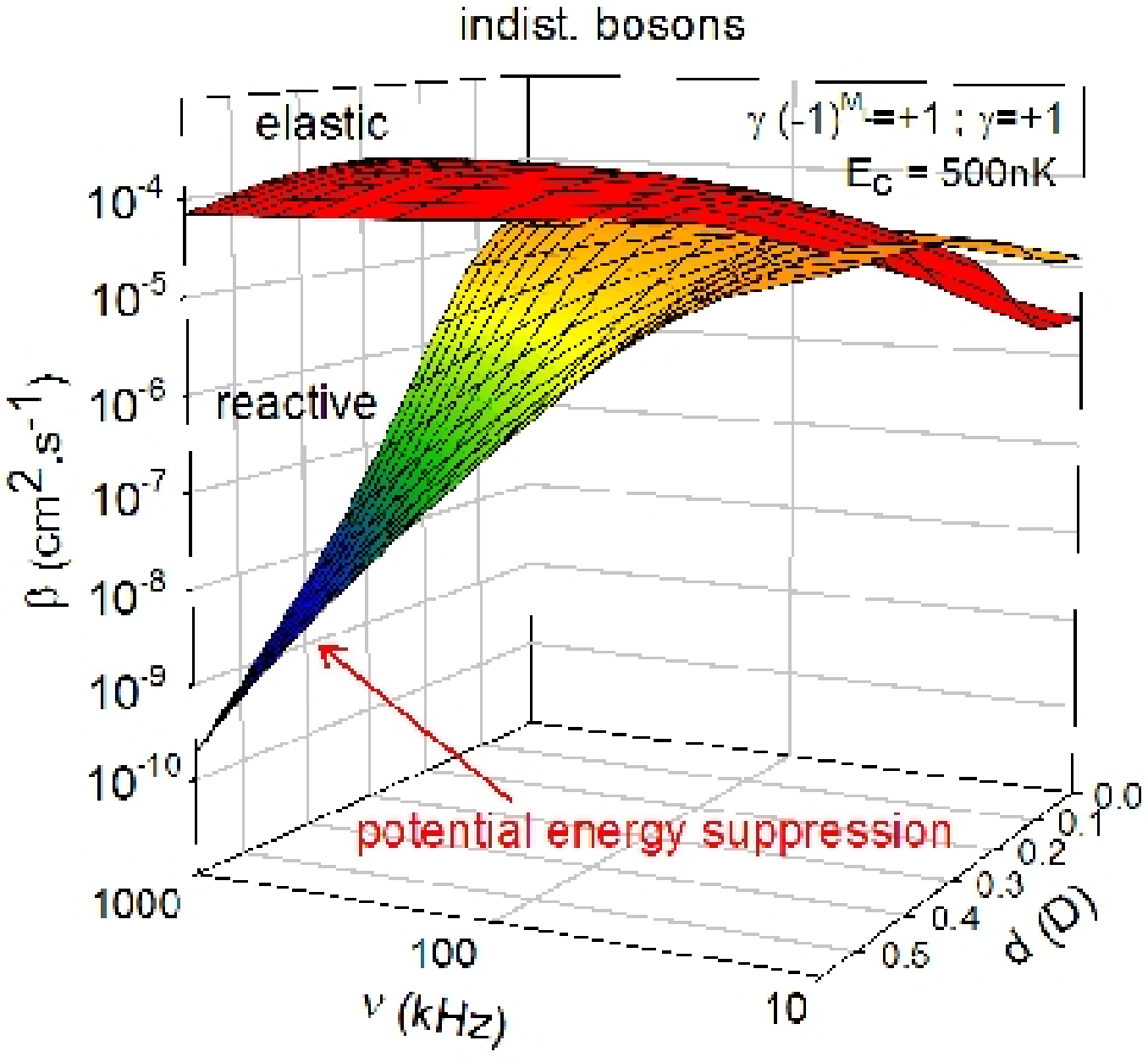}
\caption{(Color online)
Elastic and reactive rate coefficient versus $d$ and $\nu$ for indistinguishable fermions (upper panel) and
for for indistinguishable bosons (lower panel) at $E_c = 500$~nK.
The elastic curve is plotted in red.
\label{RATE-FIG}
}
\end{center}
\end{figure}

We now discuss the differences between the symmetries
rather than a specific confined collisional case.
We focus on the symmetry  $\gamma \, (-1)^{M_L} = -1$ with $\gamma=+1$ 
(black curves of the upper panel in Fig.~\ref{SPAG-FIG}) 
and on the symmetry $\gamma \, (-1)^{M_L} = +1$ with $\gamma=+1$ 
(black curves of the lower panel in Fig.~\ref{SPAG-FIG}). 
The former case corresponds to the dynamics
of identical indistinguishable fermions
and the latter to the dynamics
of identical indistinguishable bosons.
By indistinguishable, we mean identical molecules in the same internal and external states.
For the discussion, we focus only on the lowest black curve if we assume molecules
in the ground state of the trapping potential.
Two striking differences can be seen 
due to the statistics of the systems.
First, the lowest curve connects at short distance to an adiabatic curve with a $L=1$ 
adiabatic barrier $V_b$ (depicted with a green arrow) for the $\gamma \, (-1)^{M_L} = -1$ symmetry (upper panel), 
while no barrier is present ($L=0$) for the $\gamma \, (-1)^{M_L} = +1$ symmetry (lower panel).
This makes indistinguishable bosonic molecules likely to chemicaly react 
in confined geometry compared to fermionic molecules.
Second, the lowest curve ($\gamma=+1$) corresponds to ${M_L}=\pm 1$ for the first symmetry
while it corresponds to ${M_L}=0$ for the second one.
Under an electric field, the ${M_L}=0$ component always corresponds to an attractive dipole-dipole interaction
whereas the ${M_L}=1$ component corresponds to a 
repulsive dipole-dipole interaction
(which can eventually turns into an attractive one at higher dipoles~\cite{Ni10-NATURE,Quemener10-QT}).
For this rather small confinement, it means that 
we can still, up to a certain dipole, use an electric field to increase the barrier $V_b$
for indistinguishable fermions.
This is not true for indistinguishable bosons.
We will refer to this kind of suppression as ``statistical suppression'',
as it depends on the fermionic/bosonic character.
To get suppression for indistinguishable bosons, 
we will have to increase the confinement and the electric field,
which will be refered in the following to as ``potential energy suppression''. \\

To understand these two types of suppression, it is useful to  plot the 
height of the barrier $V_b$, which 
the molecules at ultralow temperature must tunnel through.
We plot this barrier in Fig.~\ref{BARRIER-FIG} for the symmetry $\gamma \, (-1)^{M_L} = -1$ with $\gamma=+1$ (upper panel)
and  for the symmetry $\gamma \, (-1)^{M_L} = +1$ with $\gamma=+1$ (lower panel),
as a function of the confinement $\nu$ and the dipole moment $d$ induced by the electric field,
for the lowest confining state.
For the first symmetry (upper panel), there are two ways to get a high barrier.
One way is for small confinements and small $d$.
The barrier increases to reach a maximum 
at $d \approx 0.15$~D.
The fact that the barrier decreases for higher dipoles comes from contributions of higher 
values of $L=3,5,...$~\cite{Ni10-NATURE,Quemener10-QT}.
For $d \approx 0.15$~D, if we follow this maximum of the three-dimensional plot for increasing confinements, 
we see that $V_b$ decreases again.
When $\nu$ increases, the zero-point energies (the ones at large $r$ in Fig.~\ref{SPAG-FIG}) increase
while the barrier is not affected at short distance because the confinement is small.
Then, the effective height of the barrier is decreased~\cite{Quemener10-2D} as $\nu$ increases.
The second way to achieve high barriers $V_b$ 
is for high dipoles and high confinements. The barrier increases monotically, emphasizing 
the electric field suppression of confined chemical rates.
When the molecules are highly confined in a two dimensional plane perpendicular to an applied electric field, 
they collide side-by-side. This repulsive electric interaction enhances the barrier and 
makes the molecules stable against collisions~\cite{Buchler07,Micheli07,Ticknor10,Quemener10-2D,Micheli10-PRL,Dincao10}.

For the second symmetry (lower panel), there is only one way to increase the barrier.
The striking difference is that 
for small confinement and/or small dipoles,
there is no barrier at short range as already seen in Fig.~\ref{SPAG-FIG}.
The only way to raise the barrier is for high confinements and high dipoles
as for the first symmetry,
where the electric dipole repulsion come 
into play.
The rise of the barrier at high confinements and high dipoles is independent
of the symmetrization of the molecules, as $V_b$ converges to similar values for both cases.  \\

The behavior of $V_b$  has crucial consequences on the dynamics of the molecules. 
To get the rate coefficients of a specific confined collision,
one has to add the rates obtained from a scattering calculation 
using the adiabatic curves of the individual 
symmetries $\gamma \, (-1)^{M_L} $ involved in the specific problem. 
The rates for the symmetry $\gamma \, (-1)^{M_L} = -1$ with $\gamma=+1$ is presented in the upper panel
and for the symmetry $\gamma \, (-1)^{M_L} = +1$ with $\gamma=+1$ in the lower panel of Fig.~\ref{RATE-FIG},
as a function of $\nu$ and $d$ for a collision energy $E_c=500$~nK.
Qualitativelly, the behavior of the reactive rates is opposite to the height of the corresponding barriers, while
the elastic rates increase only in a monotonic way with $d$ and $\nu$. 
For small confinements and dipoles (say $\nu = 20$~kHz, $d=0.15$~D), the reactive rates are suppressed
for the first symmetry (upper panel)
representing approximatelly $10^{-2}$ of the elastic rates.
No such suppression is seen for the second symmetry (lower panel). 
This shows that this statistical suppression
is only due to symmetrization requirements, but has the advantage to work 
at rather realistic experimental confinements and dipoles.
For high confinements and dipoles, the reactive rates of fermions and bosons can be suppressed by
three to four orders of magnitude 
compared to the ones at small confinements. 
This is made possible by the anisotropy of the dipolar interaction of polar molecules 
in confined geometries
as explained in Refs.~\cite{Buchler07,Micheli07,Ticknor10,Quemener10-2D,Micheli10-PRL,Dincao10}.

The elastic rates increase as $d^4$ or $d$, depending on the collision energy and magnitude of 
the dipole~\cite{Ticknor09}, and increase with $\nu$~\cite{Quemener10-2D,Micheli10-PRL}.
Therefore, this potential energy suppression of the reactive rates
and enhancement of the elastic processes
will help evaporative cooling of fermionic and bosonic molecules,
and will make amenable the creation of degenerate Fermi gases 
or Bose--Einstein condensates of polar molecules.
This suppression is not due to symmetrization requirements but to the fact that the molecules possess
a permanent electric dipole moment. Therefore, this suppression will also be effective
for molecules in distinguishable states or even for non-identical polar molecules.

It is worth noting that the fermionic statistical suppression
is still effective if the fermions are 
in different external states ($\gamma=\pm1$), since
both black and red dashed curves of the upper panel in Fig.~\ref{SPAG-FIG} 
have to be used. The red curves corresponds to a ${M_L}=0$ component,
whose barrier height $V_b$ 
decreases for increasing electric field.
There is no statistical suppression at all if the molecules are in
different internal states ($\eta=\pm1$),
because the curves from the lower panel in Fig.~\ref{SPAG-FIG} 
have to be used including the barierless curve $L=0$.
This has been confirmed experimentally~\cite{Miranda10}.

Finally, no statistical suppression can occur in the case 
of different polar molecules, for which all curves of all symmetries
in Fig.~\ref{SPAG-FIG} should be employed.
Only the potential energy suppression can apply in that case.

\section{Inelastic collisions between confining states}

\begin{figure} [t]
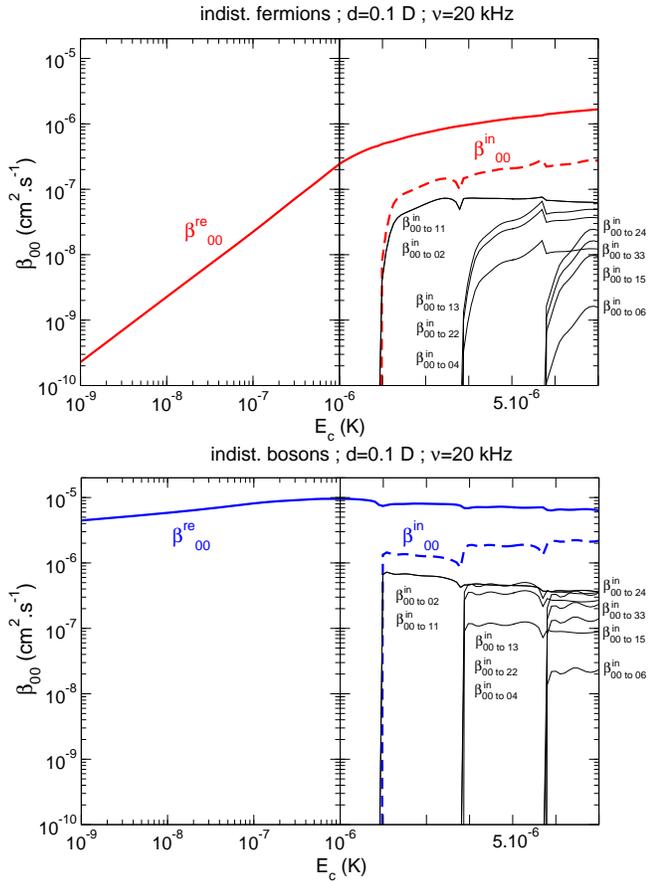

\begin{center}
\includegraphics*[width=8.5cm,keepaspectratio=true,angle=0]{figure5a.eps}  \\
\includegraphics*[width=8.5cm,keepaspectratio=true,angle=0]{figure5b.eps}  \\
\caption{(Color online) Rate coefficient $\beta^{\text{in,re}}_{00}$ versus collision energy $E_c$ for $d=0.1$~D and $\nu=20$~kHz,
for indistinguishable fermions (upper panel) and indistinguishable bosons (lower panel),
initially in the ground state of the confining trap $n_1 =  n_2 = 0$.
The thick solid (dashed) curve corresponds to reactive (inelastic) scattering.
The thin solid black lines represent the confining state-to-state rate coefficients.
\label{RATEvsEc-FIG}
}
\end{center}
\end{figure}

\begin{figure} [t]
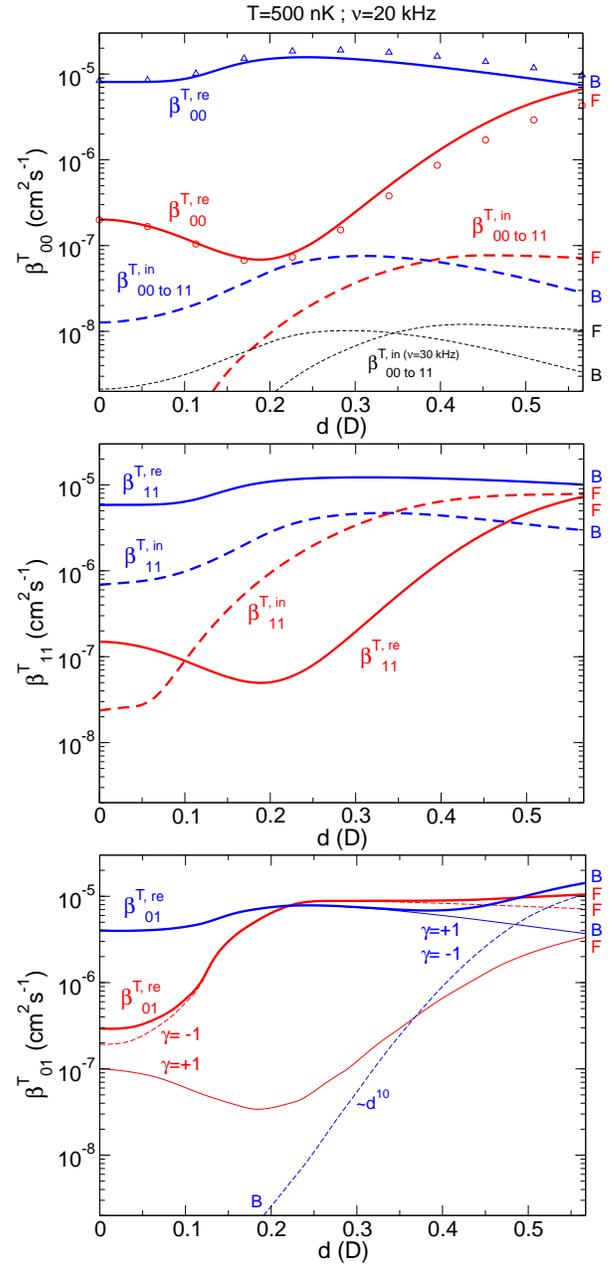

\begin{center}
\includegraphics*[width=7.8cm,keepaspectratio=true,angle=0]{figure6a.eps}  \\
\includegraphics*[width=7.8cm,keepaspectratio=true,angle=0]{figure6b.eps}   \\
\includegraphics*[width=7.8cm,keepaspectratio=true,angle=0]{figure6c.eps}
\caption{(Color online) Thermalized rate coefficient versus $d$ for $T=500$~nK and $\nu=20$~kHz.
The solid (dashed) curves correspond to reactive (inelastic) processes.
The red (blue) curves correspond to fermions (bosons) in same internal states,
but not necessarily in same external states.
The molecules are considered initially in $n_1=0, n_2=0$ (upper panel),
in $n_1=1, n_2=1$ (middle panel),
and in $n_1=0, n_2=1$ (lower panel).
\label{RATEvsT-FIG}
}
\end{center}
\end{figure}

We saw that chemical suppression of indistinguishable fermions and bosons 
can always be obtained if sufficiently high confinements and electric fields are applied.
However, the magnitude of these high confinements is still beyond of those that can be currently achieved experimentally.
For a realistic experimental frequency of $\nu \simeq 20$~kHz,
loss of indistinguishable fermions can be suppressed taking advantage of the alternative
statistical suppression whereas loss of indistinguishable bosons cannot realistically be suppressed.
Moreover, for small confinements, it is possible that higher trap confining states can be populated.
The reason is that the energy spacing between two allowed confining states
$\Delta \varepsilon = 0.96 \ \mu$K for $\nu \simeq 20$~kHz 
can be of the order of the temperature $T \simeq 500$~nK of the gas.
Then, the changing-state dynamics of molecules in small confining optical lattices 
must be understood
as well. We consider in the following fermions and bosons in same internal states
but not necessarily in the same external confining states,
for a realistic confinement of $\nu = 20$~kHz. \\

We present in Fig.~\ref{RATEvsEc-FIG} 
the non-thermalized rate coefficients $\beta^{\text{in,re}}_{00} = \sigma^{\text{in,re}}_{00} \, v$ 
as a function of the collision energy
for the inelastic and reactive processes,
for indistinguishable fermionic molecules (upper panel) and indistinguishable bosonic molecules (lower panel)
in same internal and external states.
The molecules start in $n_1=0$ and $n_2=0$ 
and $\nu = 20$~kHz, $d=0.1$~D.
States between $0 < n_1, n_2 < n^\text{max}_{\text{osc}}=7$ 
have been used for collision energy $E_c > 1$~$\mu$K
to converge these results.
At ultralow energy, the fermionic reactive rate scales as $E_c$,
and as $\ln^{-2}(\sqrt{2 \, \mu \, E_c})$ for the bosons,
in agreement with the threshold laws~\cite{Sadeghpour00,Li08}.
When the collision energy is sufficiently high, excited confining states become energetically open.
Overall, bosons react at higher rate than fermions, as expected, since
there is no barrier for bosons, whereas
there is a barrier for fermions.
Moreover, molecules that start in the ground confining state are much more 
likely to react chemically than to go to higher confining state.
The inelastic rate for the fermionic molecules is an order of magnitude
smaller than its reactive rate. It is a factor of $3-8$ smaller than the reactive rate
for the bosonic molecules.

\subsection{Gas in thermal equilibrium}

We now consider a thermal equilibrium at a temperature of $T=500$~nK.
The population $p$ of the molecules in $n_\tau$ is given by a Maxwell--Boltzmann distribution
\begin{eqnarray}
p(n_\tau) =  \frac{e^{-\frac{\varepsilon_{n_\tau}}{k_B T}}}{\sum_{n_\tau} e^{-\frac{\varepsilon_{n_\tau}}{k_B T}}}.
\label{distMB1D}
\end{eqnarray}
At $T=500$~nK in a trap with $\nu = 20$~kHz, $p(n_\tau=0) \simeq 0.852$, $p(n_\tau=1) \simeq 0.126$ and $p(n_\tau=2) \simeq 0.019$.
In the following we will neglect contribution of molecules in $n_\tau=2$,
and consider only molecules in $n_\tau=0,1$ for simplification. 
The coefficients $p(n_\tau)$ will play a role in the rate equations below. \\

We present in Fig.~\ref{RATEvsT-FIG} the thermalized rates $\beta_{00}^{T,\text{re}}$ and $\beta_{00}^{T, \text{in}}$ (upper panel),
$\beta_{11}^{T,\text{re}}$ and $\beta_{11}^{T, \text{in}}$  (middle panel), and  $\beta_{01}^{T,\text{re}}$ (lower panel),
as a function of $d$ for $\nu=20$~kHz at $T=500$~nK.
The reactive and inelastic rates are plotted as a thick and dashed solid line.
The fermionic and bosonic case are plotted in red and blue respectively.

We discuss first the case of molecules in the ground states $n_1=0, n_2=0$ (upper panel).
For bosons, the reactive rate is high
and the inelastic collision is insignificant.
For fermions however,
the inelastic rate can reach 20~\% of the amount of the reactive rate
at $d=0.23$~D. 
The magnitude of the thermalized inelastic rates is proportional to the amount of 
molecules allowed by the Maxwell-Boltzmann distribution at $T=500$~nK
to have kinetic energy greater than the first excited inelastic thresholds  $n_1=1, n_2=1$ and $n_1=0, n_2=2$
at $1.92$~$\mu$K.
We also plot in circles (fermions) and triangles (bosons) the non-thermalized reactive rate
$\beta_{00}^{\text{re}} = \sigma_{00}^{\text{re}} \, v$.
We see that $\beta_{00}^{T,\text{re}} = \beta_{00}^{\text{re}}$ is a reasonable approximation
at small dipole moments. $\beta_{00}^{\text{re}}$ differs by 35~\% from $\beta_{00}^{T,\text{re}}$
at the highest dipole, however.
This comes from the fact that at these dipoles,
the molecules do not collide in the Wigner regime anymore and the height of the barrier for fermions
(or characteristic energy for bosons) is comparable to the temperature.
Note that if the confinement is increased to $\nu=30$~kHz, 
the inelastic rate (represented as thin dashed black lines)
decreases by about an order of magnitude, because 
for a same temperature,
it is harder to excite molecules in higher confining states
as the energy thresholds increases with the confinement. 
Then the inelastic collisions for ground state molecules become less important
as the confinement increases.

If the molecules are now in the first excited states $n_1=1, n_2=1$ (middle panel),
reactive collisions, for both bosons and fermions, are about 30~\% smaller than the ones for molecules in $n_1=0, n_2=0$.
A qualitative explanation is that $n_1=1, n_2=1$ (which has a $\gamma=+1$ symmetry)
projects onto an  $n=0, N=2$ state
and a $n=2, N=0$ state (see Appendix B).
When we look at the corresponding adiabatic energies in Fig.~\ref{SPAG-FIG} for the fermions,
the $n=2$ curve connects to the $L=3$ adiabatic barrier
which is much higher than the $L=1$ barrier,
suppressing more strongly the reactive collisions
and increasing inelastic collisions.
For bosons, the reactive rates are still high compared to fermions,
because the $n=0$ curve connects to a $L=0$ curve.
However, the reactive rates are smaller than for the $n_1=0, n_2=0$ case,
because there is now the $n=2$ curve that connects to a $L=2$ curve,
suppressing chemical reactivity.
The inelastic processes are
much important in the present case
because the Maxwell-Boltzmann distribution allows all molecules
to have sufficient kinetic energy to contribute to the inelastic process, while in the precedent case, 
only a part of the molecules were allowed to contribute to the inelastic process.
For bosons, the inelastic magnitude is about half the reactive rate (at most, at $d=0.3$~D), 
but for fermions, it can even exceed the reactive rate
for $d >  0.1$~D.

Finally, we discuss the case of molecules in different states $n_1=0, n_2=1$ (lower panel).
This channel cannot decay to
the energetically allowed $n_1=0, n_2=0$ channel,
because the two channels correspond to different values of $N$.
However, the molecules are in different confining states now
so that two contributions $\gamma=\pm1$ are involved in the calculation,
and both black and red dashed curves of Fig.~\ref{SPAG-FIG} have to be used. 
This is shown in the lower panel of Fig.~\ref{RATEvsT-FIG}
as thin solid line for $\gamma=+1$ and thin dashed line for $\gamma=-1$.
Compared to fermionic molecules in the same confining states, the reactive rates are bigger.
This comes mainly from the $\gamma=-1$ contribution, which corresponds
to ${M_L}=0$ head-to-tail attractive dipolar interactions (see Tab.~\ref{TAB1}).
For bosonic molecules in different confining states, the reactive rates are similar to those for molecules in same confining states,
except that the $\gamma=-1$ contribution gives an enhancement at high dipoles due to the ${M_L}=1$
component of the $L=2$ adiabatic curve (see Tab.~\ref{TAB1}).
This component corresponds to an attractive dipolar interaction (see Eq.~8 and Eq.~9 of Ref.~\cite{Quemener10-QT})
and can enhance the reactive rate at high dipoles.
The $L=2$ barrier is high at small dipoles (see Fig.~\ref{SPAG-FIG}) and suppresses the reactive rates.
However, the strong dependence of $d^{4(L+1/2)}$ of the rates~\cite{Quemener10-QT}
leads to a $d^{10}$ dependence, as shown in the figure,
and eventually makes a significant contribution at high dipoles.

We saw on one hand that inelastic processes can be important for molecules initially in excited confining states, especially for fermions,
and that on the other hand molecules can chemically react at high rates for molecules initially in different confining states, even for fermions
because they are not indistinguishable anymore.
What are the consequences of this for the dynamics of a molecular gas?
This is what we answer in the next subsection.

\subsection{Rate equations}

The rate equations for the density of molecules $\text{n}_{n_\tau}(t)$ in state $n_\tau$  
as a function of time are given by
\begin{eqnarray}
\dot{\text{n}}_0(t) &=& - \beta_{00}^{T,\text{re}} \, \text{n}^2_0(t) -  \beta_{01}^{T,\text{re}} \, \text{n}_0(t) \, \text{n}_1(t) \nonumber \\
&-& \beta_{00 \, \text{to} \, 11}^{T} \, \text{n}^2_0(t) +  \beta_{11 \, \text{to} \, 00}^{T} \, \text{n}^2_1(t) \nonumber \\
\dot{\text{n}}_1(t) &=& - \beta_{11}^{T,\text{re}} \, \text{n}^2_1(t) -  \beta_{01}^{T,\text{re}} \, \text{n}_0(t) \, \text{n}_1(t) \nonumber \\
&-& \beta_{11 \, \text{to} \, 00}^{T} \, \text{n}^2_1(t) +  \beta_{00 \, \text{to} \, 11}^{T} \, \text{n}^2_0(t) 
\label{eqrate}
\end{eqnarray}
where $\text{n}_0(t)$ ($\text{n}_1(t)$) are the individual densities of molecules in state $n_\tau=0$ ($n_\tau=1$).
Similar equations hold for $n_\tau \ge 2$, but for simplicity, to avoid additional inelastic terms in the equations, we assumed 
$p_{n_{\tau} \ge 2} \ll p_{n_{\tau} =0,1}$.

If we assume a gas in thermal equilibrium for each time $t$, the Maxwell--Boltzmann distribution implies 
that $\text{n}_0(t)=p(0) \, \text{n}_{\text{tot}}(t)$
and  $\text{n}_1(t)=p(1) \, \text{n}_{\text{tot}}(t)$ 
(we assume  $p(0) + p(1) \simeq 1$ in our example), 
where  $\text{n}_{\text{tot}}(t)$ is the density of the total molecules.
Then by summing the equations above, we obtain the rate equation for $\text{n}_{\text{tot}}(t)$
%
%
\begin{eqnarray}
\dot{\text{n}}_\text{tot}(t) &=& - \bigg\{ p^2(0) \, \beta_{00}^{T,\text{re}} + p^2(1) \, \beta_{11}^{T,\text{re}} \nonumber \\
&+& 2 \, p(0) \, p(1) \, \beta_{01}^{T,\text{re}} \bigg \} \text{n}^2_\text{tot}(t).
\label{eqrate}
\end{eqnarray}
Inelastic rates cancel each other in the full equation,
because two molecules go back and forth in $n_1=0, n_2=0$
and $n_1=1, n_2=1$, without participating in the loss process.
Although inelastic collisions are responsible for the evolution of the individual density of molecules $\text{n}_0(t)$
and $\text{n}_1(t)$ ,
they are not responsible for the evolution
of the total density of molecules in the thermal gas.

At $T=500$~nK, $\beta_{11}^{T,\text{re}} \simeq \beta_{00}^{T,\text{re}}$ but $p^2(1) \ll p^2(0)$
so that the second term on the right hand side of the equation above can be neglected.
As a result the density of the total molecules
will show 
a faster decay 
due to a fast rate  $2 \, p(0) \, p(1) \, \beta_{01}^{T,\text{re}}$
and a slow decay due to a slow rate $p^2(0) \, \beta_{00}^{T,\text{re}}$.
For example for fermionic KRb at $d=0.2$~D, 
$2 \, p(0) \, p(1) \, \beta_{01}^{T,\text{re}} \simeq  1.4 \, 10^{-6}$~cm$^2$~s$^{-1}$
and $p^2(0) \, \beta_{00}^{T,\text{re}} \simeq 5. 10^{-8}$~cm$^2$~s$^{-1}$.
The fast and slow decays are 
due to high inter-states reactive rates (collisions between different confining states)
and low intra-states reactive states (collisions between same confining states).
The two types of decay can be tuned by changing the relative populations $p(0)$ and $p(1)$,
by changing the temperature $T$ and/or the confinement $\nu$.
Note that even if the population of the molecules in different confining states are not given by a
Maxwell--Boltzmann distribution, say for example $p(0) = 0.5 $ and $p(1) = 0.5 $,
and is independent of time,
inelastic rates still cancel each other in the equation for the total density of molecules.
Again, inelastic collisions play a role in the loss of molecules from individual trap levels,
but do not for the loss of the total molecules.
These theoretical findings well support recent experimental data of 
confined fermionic KRb molecules in electric fields~\cite{Miranda10}.

\section{Conclusion}

We have developed in detail a rigorous time-independent quantum 
formalism to describe the dynamics of particles
with permanent electric dipole moments in a confined geometry,
by treating the reactive chemistry using an absorbing potential. 
Elastic, reactive and inelastic rate coefficients can be computed
for a given collision energy, temperature, confinement and dipole moment (or electric field),
for a system of fermionic or bosonic molecules.
The selection rules play an important role for the dynamics of confined molecules
and have dramatic effects on the collisional properties.
Different rates are obtained 
for fermionic/bosonic molecules in same/different confining states.
Two kinds of suppression can occur for chemical reactions:
a statistical suppression applies only for fermions
at rather small induced dipoles and confinements realistically accessible
in an experiment,
and a potential energy suppression
applies for both fermions and bosons at rather high induced dipoles and confinements.
Inelastic rates can be important, even as high as reactive rates
for molecules initially in excited states.
However, the inelastic rates do no play a role in the 
loss process of the total number of molecules in a gas,
since molecules are inelastically excited and relaxed, back and forth.
Only reactive rates are responsible for the evolution of the loss of the total molecules.
Fast and slow decays of the molecules can be seen due to inter-state and intra-state confined collisions.
This work has been highly motivated by recent experiments
of KRb molecules in confined geometry and electric field,
and has proved very good theoretical support
for the experimental observations~\cite{Miranda10}.

\section*{Acknowledgments}

This material is based upon work supported by the Air Force Office of Scientific Research
under the Multidisciplinary University Research Initiative Grant No. FA9550-09-1-0588.
We also acknowledge the financial support of the National Institute of Standards and Technolgy
and the National Science Foundation.
We thank 
M. H. G. de Miranda, A. Chotia, B. Neyenhuis, D. Wang, S. Ospelkaus, S. Moses,
D. S. Jin and J. Ye for stimulating discussions
about the KRb experiment.

\section*{Appendix A: Relation between $|n_1 \, n_2 \rangle $ and $|n , N \rangle $}

In Eq.~\eqref{gNgntogn1gn2}, we use the following characteristics~\cite{Wolfram}
%
%
\begin{multline}
g_{n_\tau}(x) = \sqrt{ \frac{1}{2^{n_\tau} \, {n_\tau}!} } \, \left( \frac{m_\tau \, \omega}{\pi \, \hbar} \right)^{1/4} \\
\, e^{ - \frac{m \, \omega \, x^2}{2 \, \hbar}} \, H_{n_\tau}(\sqrt{m \, \omega / \hbar} \, x) 
\end{multline}
\begin{multline}
H_{n_\tau}(x+y) = 2^{-{n_\tau}/2} \, \sum_{k=0}^{n} \frac{{n_\tau}!}{k! ({n_\tau}-k)!} \\
 H_{k}(x \, \sqrt{2}) \,  H_{{n_\tau}-k}(y \, \sqrt{2}) 
\end{multline}
\begin{multline}
H_{n_\tau}(x) \, H_{m_\tau}(x) = \sum_{k=0}^{\text{min}({n_\tau},{m_\tau})} \\
\frac{{m_\tau}!}{k! ({m_\tau}-k)!} \, \frac{{n_\tau}!}{k! ({n_\tau}-k)!}
\, H_{-2k+{m_\tau}+{n_\tau}}(x) \, 2^{k} \, k!.
\end{multline}

The individual $|n_1 \, n_2 \rangle $ states are written in terms of the relative/CM $|n , N \rangle $ states by
\begin{eqnarray}
| 0 0 \rangle  &=& | 0, 0 \rangle   \nonumber \\
| 0 1 \rangle  &=& \frac{1}{\sqrt{2}}  | 0, 1 \rangle  + \frac{1}{\sqrt{2}}  | 1, 0 \rangle   \nonumber  \\
| 1 0 \rangle  &=& \frac{1}{\sqrt{2}}  | 0, 1 \rangle  - \frac{1}{\sqrt{2}}  | 1, 0 \rangle   \nonumber  \\
| 0 2 \rangle  &=& \frac{1}{{2}}  | 0, 2 \rangle + \frac{1}{\sqrt{2}}  | 1, 1 \rangle  + \frac{1}{{2}}  | 2, 0 \rangle  \nonumber   \\
| 2 0 \rangle  &=& \frac{1}{{2}}  | 0, 2 \rangle - \frac{1}{\sqrt{2}}  | 1, 1 \rangle  + \frac{1}{{2}}  | 2, 0 \rangle   \nonumber  \\
| 1 1 \rangle  &=& \frac{1}{\sqrt{2}}  | 0, 2 \rangle  - \frac{1}{\sqrt{2}}  | 2, 0 \rangle  . 
\end{eqnarray}

\section*{Appendix B: Relation between $|n_1 \, n_2 , \gamma \rangle $ and $|n , N \rangle $}

Using Eq.~\eqref{symstaten1n2} and Appendix A,
the symmetrized individual $|n_1 \, n_2, \gamma \rangle $ states 
are written in terms of the relative/CM $|n , N \rangle $ states by
\begin{eqnarray}
| 0 0 , \gamma=+1 \rangle  &=& | 0, 0 \rangle  \nonumber \\
| 0 1 , \gamma=+1  \rangle  &=& | 0, 1 \rangle  \nonumber  \\ 
| 0 2 , \gamma=+1  \rangle  &=& \frac{1}{\sqrt{2}}  | 0, 2 \rangle  + \frac{1}{\sqrt{2}}  | 2, 0 \rangle  \nonumber   \\
| 1 1 , \gamma=+1  \rangle  &=& \frac{1}{\sqrt{2}}  | 0, 2 \rangle  - \frac{1}{\sqrt{2}}  | 2, 0 \rangle  \nonumber   \\
| 1 2 , \gamma=+1  \rangle  &=& \sqrt{\frac{12}{16}}  | 0, 3 \rangle  - \frac{1}{2} | 2, 1 \rangle  \nonumber \\ 
| 2 2 , \gamma=+1  \rangle  &=& \sqrt{\frac{3}{8}}  | 0, 4 \rangle  - \frac{1}{2} | 2, 2 \rangle + \sqrt{\frac{3}{8}} | 4, 0 \rangle  \nonumber  \\
| 0 1 , \gamma=-1  \rangle  &=& | 1, 0 \rangle  \nonumber  \\
| 0 2 , \gamma=-1  \rangle  &=& | 1, 1 \rangle   \nonumber \\
| 1 2 , \gamma=-1  \rangle  &=& \frac{1}{2} | 1, 2 \rangle  - \sqrt{\frac{12}{16}} | 3, 0 \rangle   .
\end{eqnarray}
Note that $(-1)^{n_1 + n_2} = (-1)^{n + N}$.

\begin{table}[h]
\begin{center}
\begin{tabular}{|c | c | c | c | c | c |}
\hline
fermions &  $\eta$ & $L$ & $\gamma$ & ${M_L}$ & $n$   \\ [0.5ex]
\hline
 & +1 & 1,3,5 ... & +1  &  1,3,5 ... & 0,2,4 ... \\
 &    & 1,3,5 ... & -1  &  0,2,4 ... & 1,3,5 ... \\
 & -1 & 0,2,4 ... & +1  &  0,2,4 ... & 0,2,4 ... \\
 &    & 2,4,6 ... & -1  &  1,3,5 ... & 1,3,5 ... \\ [0.5ex]
\hline
bosons & $\eta$ & $L$ & $\gamma$ & ${M_L}$ & $n$   \\ [0.5ex]
\hline
 & +1 & 0,2,4 ... & +1  & 0,2,4 ... & 0,2,4 ... \\
 &    & 2,4,6 ... & -1  & 1,3,5 ... & 1,3,5 ... \\
 & -1 & 1,3,5 ... & +1  & 1,3,5  ...& 0,2,4 ...  \\
 &    & 1,3,5 ... & -1  & 0,2,4  ...& 1,3,5 ...  \\
\hline
\end{tabular}
\end{center}
\caption{Selection rules for
the dynamics of identical bosons and fermions
in confined two dimensional geometry.
\label{TAB1}
}
\end{table}

\section*{Appendix C: Selection rules}

For initial states $n_1,n_2$
and final states $n_1',n_2'$,
since components of different $N$ do not mix together
in the collision process, we have 
\begin{eqnarray}
(-1)^{n_1 + n_2} = (-1)^{n_1' + n_2'} 
\end{eqnarray}
after a collision.

At long range, in cylindrical coordinates, if we use the symmetrized individual representation $|n_1 \, n_2 , \gamma \rangle$,
the permutation $P$ requires the substitutions $z_1 \to z_2, z_2 \to z_1, \varphi \to \varphi + \pi$
which leads to the selection rule
\begin{eqnarray}
\eta \, \gamma \, (-1)^{M_L} = \epsilon_P .
\end{eqnarray}
If we use the relative representation $|n, \, N \rangle$ states,
then the permutation $P$ requires the substitutions $z \to -z, \varphi \to \varphi + \pi$
which leads to
\begin{eqnarray}
\eta \, (-1)^{M_L+n} = \epsilon_P ,
\end{eqnarray}
from the properties of the $g_n(z)$ functions.
At short range, in spherical coordinates, using the Legendre polynomials,
the permutation $P$ requires the substitutions $\theta \to \pi - \theta, \varphi \to \varphi + \pi$
which leads to
\begin{eqnarray}
\eta \, (-1)^L = \epsilon_P . 
\end{eqnarray}
We summarize in Tab.~\ref{TAB1} the different selection rules for identical
bosons and fermions.


\begin{thebibliography}{27}


\bibitem{Ospelkaus10-SCIENCE}
S. Ospelkaus, K.-K. Ni, D. Wang, M. H. G. de Miranda, B. Neyenhuis, G. Qu\'em\'ener,
P. S. Julienne, J. L. Bohn, D. S. Jin, and J. Ye,
Science {\bf 327}, 853 (2010).

\bibitem{Ni10-NATURE}
K.-K. Ni, S. Ospelkaus, D. Wang, G. Qu\'em\'ener, B. Neyenhuis, M. H. G. de Miranda,
J. L. Bohn, J. Ye, and D. S. Jin,
Nature,  {\bf 464} 1324 (2010).

\bibitem{Ni08-SCIENCE}
K.-K. Ni, S. Ospelkaus,
M. H. G. de Miranda, A. Pe'er, B. Neyenhuis, J. J. Zirbel, S. Kotochigova,
P. S. Julienne, D. S. Jin, and J. Ye,
Science {\bf 322}, 231 (2008).

\bibitem{Ospelkaus10-PRL}
S. Ospelkaus, K.-K. Ni, G. Qu\'em\'ener, B. Neyenhuis,
D. Wang, M. H. G. de Miranda, J. L. Bohn, J. Ye, and D. S. Jin,
Phys. Rev. Lett. {\bf 104}, 030402 (2010).

\bibitem{Aikawa10-ARXIV}
K. Aikawa, D. Akamatsu, M. Hayashi, K. Oasa, J. Kobayashi, P. Naidon, T. Kishimoto, M. Ueda, and S. Inouye, 
accepted to Phys. Rev. Lett.,
arXiv:1008.5034.

\bibitem{Sage05}
J. M. Sage,
S. Sainis, T. Bergeman, and D. DeMille,
Phys. Rev. Lett. {\bf 94}, 203001 (2005).

\bibitem{Deiglmayr08}
J. Deiglmayr,
A. Grochola, M. Repp, K. M\"ortlbauer,
C. Gl\"uck, J. Lange, O. Dulieu, R. Wester, and M. Weidem\"uller,
Phys. Rev. Lett. {\bf 101}, 133004 (2008).

\bibitem{Zuchowski10}
P. S. $\dot{\text{Z}}$uchowski and J. M. Hutson,
Phys. Rev. A {\bf 81}, 060703(R) (2010).

\bibitem{Byrd10}
J. N. Byrd, J. A. Montgomery Jr., and R. C\^ot\'e,
Phys. Rev. A {\bf 82}, 010502(R) (2010).

\bibitem{Meyer10}
E. R. Meyer and J. L. Bohn,
accepted to Phys. Rev. A,
arXiv:1004.3317.

\bibitem{Quemener10-QT}
G. Qu\'em\'ener and J. L. Bohn,
Phys. Rev. A {\bf 81}, 022702 (2010).

\bibitem{Buchler07}
H. P. B\"uchler, E. Demler, M. Lukin, A. Micheli, N. Prokofiev, G. Pupillo, and P. Zoller,
Phys. Rev. Lett. {\bf 98}, 060404 (2007).

\bibitem{Micheli07}
A. Micheli, G. Pupillo, H. P. B\"uchler, and P. Zoller,
Phys. Rev. A {\bf 76}, 043604 (2007).

\bibitem{Ticknor10}
C. Ticknor,
Phys. Rev. A {\bf 81}, 042708 (2010).

\bibitem{Quemener10-2D}
G. Qu\'em\'ener and J. L. Bohn,
Phys. Rev. A {\bf 81}, 060701(R) (2010).

\bibitem{Micheli10-PRL}
A. Micheli, Z. Idziaszek, G. Pupillo, M. A. Baranov, P. Zoller, and P. S. Julienne
Phys. Rev. Lett. {\bf 105}, 073202 (2010).

\bibitem{Dincao10}
J. P. D'Incao and C. H. Greene,
in preparation.

\bibitem{Carr09}
L. D. Carr, D. DeMille, R. V. Krems, and J. Ye,
New J. Phys. {\bf 11}, 055049 (2009).

\bibitem{Micheli06-NATURE}
A. Micheli, G. K. Brennen and P. Zoller,
Nat. Phys. {\bf 2}, 341 (2006).

\bibitem{Demille02}
D. DeMille,
Phys. Rev. Lett. {\bf 88}, 067901 (2002).

\bibitem{Yelin06}
S. F. Yelin, K. Kirby, and R. C\^ot\'e,
Phys. Rev. A {\bf 74}, 050301(R) (2006).

\bibitem{Miranda10}
M. H. G. de Miranda, A. Chotia, B. Neyenhuis, D. Wang, G. Qu\'em\'ener, S. Ospelkaus,
J. L. Bohn, J. Ye, and D. S. Jin, submitted.

\bibitem{Petrov01}
D. S. Petrov and G. V. Shlyapnikov,
Phys. Rev. A {\bf 64}, 012706 (2001).

\bibitem{Li09}
Z. Li and R. V. Krems,
Phys. Rev. A {\bf 79}, 050701(R) (2009).

\bibitem{Granger04}
B. E. Granger and D. Blume,
Phys. Rev. Lett. {\bf 92}, 133202 (2004).

\bibitem{Omahony91}
P. F. O'Mahony and F. Mota-Furtado, 
Phys. Rev. Lett. {\bf 67},  2283 (1991).

\bibitem{Soldan10}
P. Sold\'an,
Phys. Rev. A {\bf 82}, 034701 (2010).

\bibitem{Pack87}
R. T Pack and G. A. Parker,
J. Chem. Phys. {\bf 87}, 3888 (1987).

\bibitem{Launay89}
J.-M. Launay and M. Le Dourneuf,
Chem. Phys. Lett. {\bf 163}, 178 (1989).

\bibitem{Quemener06-PHD}
G. Qu\'em\'ener, PhD Thesis, University of Rennes, France (2006),
http://tel.archives-ouvertes.fr/tel-00204105.

\bibitem{Idziaszek10-PRL}
Z. Idziaszek and P. S. Julienne
Phys. Rev. Lett. {\bf 104}, 113202 (2010).  

\bibitem{Idziaszek10-PRA}
Z. Idziaszek, G. Qu\'em\'ener, J. L. Bohn and P. S. Julienne,
Phys. Rev. A {\bf 82}, 020703(R) (2010).

\bibitem{Johnson73}
B. R. Johnson, J. Comp. Phys. {\bf 13}, 445 (1973).

\bibitem{Lapidus82}
I. Richard Lapidus,
Am. J. Phys. {\bf 50}, 45 (1982).

\bibitem{Adhikari86}
S. K. Adhikari,
Am. J. Phys. {\bf 54}, 362 (1986).

\bibitem{Naidon06}
P. Naidon and P. S. Julienne,
Phys. Rev. A {\bf 74}, 062713 (2006).

\bibitem{Burke99}
J. P. Burke, Jr., PhD Thesis, University of Colorado (1999),
http://jilawww.colorado.edu/pubs/thesis/burke.

\bibitem{AS}
M. Abramowitz and I. A. Stegun,
{\it Handbook of mathematical functions}, Dover editions.

\bibitem{Ticknor09}
C. Ticknor,
Phys. Rev. A {\bf 80}, 052702 (2009).

\bibitem{Sadeghpour00}
H. R. Sadeghpour, J. L. Bohn, M. J. Cavagnero, B. D. Esry, I. I. Fabrikant,
J. H. Macek, and A. R. P. Rau,
J. Phys. B: At. Mol. Opt. Phys. {\bf 33}, R93 (2000).

\bibitem{Li08}
Z. Li, S. V. Alyabyshev, and R. V. Krems,
Phys. Rev. Lett. {\bf 100}, 073202 (2008).

\bibitem{Wolfram}
http://functions.wolfram.com

\end{thebibliography}
\end{document}